\documentclass[11pt]{article}
\usepackage{epsfig}
\usepackage{graphicx}
\usepackage{amsmath}
\usepackage{hhline}
\usepackage{amssymb}
\usepackage{times}
\usepackage{cite}
\usepackage{array}
\usepackage{multirow}

\usepackage{color}
\definecolor{rltred}{rgb}{0.75,0,0}
\definecolor{rltgreen}{rgb}{0,0.5,0}
\definecolor{rltblue}{rgb}{0,0,0.75}

\newlength{\dinwidth}
\newlength{\dinmargin}
\begin{document}
\begin{titlepage}

\noindent
Date:        \today       \\
                
\vspace{2cm}

\begin{center}
\begin{Large}

{\bf Behavior of $\sigma^{\gamma P}$ at Large Coherence Lengths}

\vspace{2cm}

Allen Caldwell \\
Max Planck Institute for Physics (Werner-Heisenberg-institut) \\
Munich, Germany

\end{Large}
\end{center}

\vspace{2cm}

\begin{abstract}

Various parametrizations of $\sigma^{\gamma P}$ are tried out on the small-$x$ fixed target and HERA data. A two-Pomeron type parametrization is found to give the best reproduction of the data.  The data indicate that the value of $\lambda_{\rm eff}$ for parametrizations of the form $\sigma^{\gamma P}|_{Q^2} \propto l^{\lambda_{\rm eff}}$ approaches a constant value at high $Q^2$ .  The extrapolated values of $\sigma^{\gamma P}$ to very long coherence lengths are found to cross in some parametrizations for $l\geq 10^7$~fm, indicating the possibility that $\sigma^{\gamma P}$ becomes $Q^2$ independent at large values of the coherence length $l$.

\end{abstract}
\end{titlepage}

\section{Introduction}

The small-$x$ behavior of the proton structure function $F_2$ is striking and has inspired many
models and parametrizations.  In this paper, fits using different parametrizations of the photon-proton cross section  are compared.  The standard Hand convention~\cite{ref:Hand} is used to define the photon flux, yielding the relation:
$$F_2^P=\frac{Q^4(1-x)}{4\pi^2\alpha (Q^2+(2xM_P)^2)}\sigma^{\gamma P} $$
where $\alpha$ is the fine structure constant and $M_P$ is the proton mass.
Given a parametrization for $\sigma^{\gamma P} $, we can compare predicted values of $F_2^P$ to data.  Data from E665, NMC, H1 and ZEUS have been used for these fits. 

\begin{figure}[hbpt]
\begin{center}
   \includegraphics[width=0.5\textwidth]{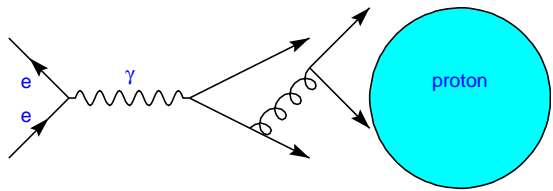}
\caption{\it Photon fluctuations scattering on the proton in the proton rest frame.}
\label{fig:proton}
\end{center}
\end{figure}

The behavior of $\sigma^{\gamma P}$ is studied in the proton rest frame in terms of the coherence length of the photon fluctuations, $l$, and the virtuality, $Q^2$.  The physics picture is given in Fig.~\ref{fig:proton}, where the electron acts as a source of photons, which in turn acts as a source of quarks, antiquarks and gluons.   The partonic wavefunction of the photon state is dependent on $l$ and $Q^2$. The proton is viewed as a set of interaction centers for the incoming partons. 

At small-$x$, the coherence length of the photon fluctuations~\cite{ref:LES} in the proton rest frame is a more intuitive variable than Bjorken-$x$, and we will use this variable and $Q^2$ in our parametrizations.  Recall the definition of the coherence length, $l$:
$$l=\frac{\hbar c}{\Delta E}$$
where $\Delta E$, the change in energy of the photon as it fluctuates into a system of quarks and gluons, is given by
\begin{eqnarray}
\label{eq:cohere}
\Delta E & \approx & \frac{M^2+Q^2}{2\nu} \\
               & \approx & \frac{Q^2}{\nu} 
\end{eqnarray}
where we have set the mass of the hadronic system into which the photon fluctuated, $M=Q$, and $\nu$ is the photon energy. This yields
$$l\equiv \frac{\hbar c}{2x M_P}\;\; .$$  
Note that $M_P$ only appears because of the definition of $x$.  This is the standard expression used in the literature for the coherence length, although other possibilities have also been discussed taking different choices for $M^2$~\cite{ref:schildknecht}.

  The data used in this analysis are restricted to small $x$ values so that $l$ is many times larger than the proton radius.  For compact photon fluctuations, the maximum value of $\sigma^{\gamma P}$ is given by the size of the proton multiplied by $\alpha$, giving roughly $200$~$\mu$barn.  However, pQCD calculations have pointed to the property of `color transparency' for small dipoles~\cite{ref:transparency}, indicating that at large $Q^2$ the cross section should behave as $\sigma^{\gamma P} \propto 1/Q^2$.  I.e., the proton appears almost transparent for small dipoles with the cross section proportional to the size of the photon.  The photon state will have a maximum size which is expected to be set by the mass of the lightest vector meson.  An effective mass will be used as a free parameter in the fits.  All parametrization will therefore have a basic term:
$$\sigma_0 \frac{M^2}{Q^2+M^2}$$
where $\sigma_0$ is expected to be a typical hadronic cross section (multiplied by $\alpha$).
 The parametrizations will  primarily be used to discriminate the $l$ dependence of the photon-proton cross section.
 
 In the next sections, we start by reviewing the data sets which have been used for the analysis.  This is followed by a discussion of the parametrizations which have been studied, and a description of the fitting technique employed (Bayesian analysis based on Markov Chains).  The results are then presented and discussed.

\section{Data Sets}
\label{sec:data}
The proton DIS data used in this analysis were taken from the Durham Data Base\footnote{http://durpdg.dur.ac.uk/HEPDATA/}, with the exception of the H1 data from the 1999-2000 running period which was taken from the H1 collaboration web page.  For the final results, only data with $x<0.025$ have been used.  Data from the fixed target experiments NMC and E665 were used, as well as data from H1 and ZEUS.  BCDMS data was not included since only a few data points are available from this experiment in the $x$ range of interest.   Table~\ref{tab:Data} summarizes the main aspects of the data sets used in this analysis.

\begin{table}[htdp]
\caption{\it The data sets used in this analysis.}
\begin{center}
\begin{tabular}{|c|cccccc|c|}
\hline
Experiment & $x_{\rm min}$ & $Q^2_{\rm min}$ & $Q^2_{\rm max}$ & & \# data & & Ref.\\
 &  & (GeV$^2$) &  (GeV$^2$) & $x<0.025$ & $x<0.02$ & $x<0.01$ & \\
\hline
E665 & $8.9 \cdot 10^{-4}$ & $0.229$  & $13.4$ & 64 & 58 &45 &\cite{ref:E665}\\
NMC & $3.5 \cdot 10^{-3}$ & $0.8$ &  $7.2$ & 51 &39 &13 & \cite{ref:NMC}\\
H1 SVTX & $6.1 \cdot 10^{-6}$ & $0.35$  & $3.5$ & 44 & 44 &44 &\cite{ref:H1SVTX}\\
ZEUS 97 & $6.3 \cdot 10^{-5}$ & $2.7$  & $800$ & 147 & 131 &116 &\cite{ref:ZEUS97}\\
H1 97 & $3.2 \cdot 10^{-5}$ & $1.5$  & $120$ & 116 & 110 &100 &\cite{ref:H197}\\
ZEUS BPT & $6.2 \cdot 10^{-7}$ & $0.045$  & $0.65$ & 70 & 70 &70 &\cite{ref:ZEUSBPT}\\
H1 hiQ & $3.0 \cdot 10^{-3}$ & $150$  & $1000$ & 26 & 18 &11 &\cite{ref:H1hiQ}\\
H1 9900 & $2.0 \cdot 10^{-3}$ & $100$  & $1000$ & 27 & 19 & 13 &\cite{ref:H19900}\\
\hline 
 \end{tabular}
\end{center}
\label{tab:Data}
\end{table}%

In the experimental analyses, $F_2$ is extracted by first correcting the observed data for several effects, including QED radiative effects and the contribution from the longitudinal structure function, $F_L$.  Different codes and assumptions have been used for these purposes by  the different collaborations.  In addition, the QED radiative corrections depend strongly on the technique used to measure the kinematic variables, and are usually largest at large $y=Q^2/s$.  The correction for $F_L$ is proportional to $y^2$, and therefore also has the largest impact at large $y$.  On the other hand, energy scale uncertainties have the biggest impact at small $y$, and give uncertainties which increase strongly at small $y$.  These effects result in correlated variations of the data.  To account for this, the data were multiplied in some fits by a function 
\begin{equation}
\label{eq:systs}
f(y)=1+\frac{a}{y}+by^2
\end{equation}
where $a$ and $b$ are fit parameters.  The parameter values were assumed to be Gaussian distributed with mean 0.  The Gaussian widths  for the variations are given in Table~\ref{tab:systs}.  

Table~\ref{tab:systs} also lists the normalization uncertainty on the different data sets.  Since the fits described in the next section included an overall constant, $\sigma_0$, the normalization of one of the data sets was fixed - the ZEUS97 data set.  The normalization uncertainties are therefore relative to the normalization of the ZEUS 97 data.  For this reason, the normalization uncertainty of the ZEUS BPT data set used in the fits was only $1$~\%, since the same running period and the same luminosity measurement was used as for the ZEUS97 data set.  The remaining normalization uncertainty accounts for the efficiency of the BPT system itself.  The NMC data at muon beam energies of $90,120,200,280$~GeV as reported in reference~\cite{ref:NMC} were combined into one file and an overall $2.5$~\% normalization uncertainty was assigned (i.e., the relative normalization uncertainty of 2~\% discussed in \cite{ref:NMC} was ignored).   The H1 data were each assigned an independent normalization factor since they corresponded to different data taking periods or rather different experimental conditions.

In cases where the correlated systematic uncertainties as a function of $y$ were used, the uncertainties on individual data points were calculated as follows:
\begin{equation}
\label{eq:uncorr}
e_i^2=stat_i^2+uncorr_i^2
\end{equation}
where $stat_i$ is the statistical uncertainty on point $i$ and $uncorr_i$ is the uncorrelated systematic uncertainty.  The latter was calculated as
$$uncorr_i^2=max\left[syst_i^2-corr_i^2,0\right]$$
where $syst_i$ is the total systematic uncertainty reported on point $i$ (not including the overall normalization uncertainty) and $corr_i$ is given by
\begin{equation}
\label{eq:corr}
corr_i^2=\left[\left(\frac{\sigma_a}{y}\right)^2+(\sigma_by^2)^2\right]F_2^2 \;\; .
\end{equation}

In the fits where the $y$-dependent systematic uncertainties were not used (the majority of fits), then the uncertainty on each point was simply taken as the sum in quadrature of the statistical and systematic uncertainties:
\begin{equation}
\label{eq:quad}
e_i^2=stat_i^2+syst_i^2 \;\; .
\end{equation}

The systematic effects were then studied by changing the kinematic range over which the fits were performed.  For all fits, the relative normalization of the different data sets was allowed to vary.

\begin{table}[htdp]
\caption{\it The systematic uncertainties used for the different data sets.  Note that the normalization uncertainty of the ZEUS97 data was set to 0 in the fits.  The normalization uncertainty of the ZEUS BPT data was reduced to $1$~\% since the same luminosity measurement was used as for the ZEUS97 data.}
\begin{center}
\begin{tabular}{|c|ccc|}
\hline
Experiment & Normalization  & $\sigma_a$ & $\sigma_b$ \\
                     & Uncertainty (\%) & & \\
\hline
E665 & $1.8$ & $0.005$ & $0.15$ \\
NMC & $2.5$ & $0.001$ & $0.2$ \\
H1 SVTX & $3.0$ & $0.0003$ & $0.2$ \\
ZEUS 97 &$2.2 (0)$ & $0.0003$ & $0.2$ \\
H1 97 & $1.7$ & $0.0005$ & $0.2$ \\
ZEUS BPT & $2.3 (1)$ & $0.0002$ & $0.1$ \\
H1 hiQ & $2.0$ & $0.0005$ & $0.1$ \\
H1 9900 & $1.8$ & $0.0005$ & $0.1$ \\
\hline 
 \end{tabular}
\end{center}
\label{tab:systs}
\end{table}%

\section{Parametrizations}
\label{sec:params}

As mentioned in the introduction, all parametrizations studied have the same basic $Q^2$ dependence, and are of the form
$$\sigma^{\gamma P}=\sigma_0 \frac{M^2}{Q^2+M^2} f(l) \;\; .$$
They are distinguished by the form chosen for $f(l)$.  The various forms attempted were:
\paragraph{D}
A form inspired by the observed features of the HERA data~\cite{ref:ZEUSpheno}.
$$f(l)=\left[ \frac{l}{l_0} \right]^{\lambda(Q^2)}$$ where
$\lambda={\epsilon_0+\epsilon' \ln(Q^2+Q^2_0)} $
giving the 6 free model parameters: $\sigma_0,M^2,l_0,\epsilon_0,\epsilon',Q^2_0$.
In this parametrization there is a `soft' energy dependence at small $Q^2$, $\sigma\propto l^{\epsilon_0}$, with a linear increase of the power of $l$ with  $\ln Q^2$ at high $Q^2$. The variation of $\lambda$ with $Q^2$ is a well-known feature of the HERA data, so it is included here explicitly.  The expected values of the parameters are:
\begin{itemize}
\item[$\sigma_0$]  is expected to be of order $100$~{$\mu$}barn, since this is the order of magnitude of the measured photoproduction cross section.  
\item[$M^2$] is expected to be around $M_{\rho}^2\approx 0.6$~GeV$^2$, since the $\rho$ is the lightest vector meson;
\item[$l_0$] could be of order  of the proton radius ($0.9$~fm), as this is the only dimensionful scale available.  Another guide is the minimum value of $l\approx 0.1$~fm;
\item[$\epsilon_0$] should be very close to the `soft Pomeron' value of Donnachie and Landshoff~\cite{ref:DL}, and the more recent update~\cite{ref:cudell} which yields about $0.1$;
\item[$\epsilon'$] gives the rate at which $\lambda$ is increasing with $\ln Q^2$ at high $Q^2$.  The data indicates that the value is around $0.05$;
\item[$Q^2_0$] is presumably near $1$~GeV$^2$, since this is where the rise in the energy dependence has been observed.
\end{itemize}
The prior probability distributions are assumed flat within large ranges around the expected values.  All parameters are constrained to be positive.

\paragraph{2P} A two-Pomeron model~\cite{ref:DL2} inspired parametrization:
$$f(l)=\left[ \frac{l}{l_0} \right]^{\epsilon_0+(\epsilon_1-\epsilon_0) \sqrt{\frac{Q^2}{Q^2+\Lambda^2}}} $$
giving the 6 free parameters: $\sigma_0,M^2,l_0,\epsilon_0,\epsilon_1,\Lambda^2$.
In this parametrization, there is a smooth transition from a `soft Pomeron' with intercept $1+\epsilon_0$ to a `hard Pomeron' with intercept $1+\epsilon_1$. The square-root is not obvious, but was found to give excellent fits.  The first four model parameters are as in the D model above, and should have similar values in the fits.  For the other parameters:
\begin{itemize}
\item[$\epsilon_1$] a value in the range of $0.3-0.5$ is expected.  It should clearly be larger than the largest values of $\lambda_{\rm eff}$ which have been seen in previous fits to HERA data, but is not expected to be larger than the LO BFKL Pomeron~\cite{ref:BFKL} value of $0.5$;
\item[$\Lambda^2$] gives an intermediate $Q^2$ value where the `soft' and `hard' Pomerons are both contributing.  It is clearly much larger than $1$~GeV$^2$.  In the fits, values up to $100$~GeV$^2$ were allowed. 
\end{itemize}

\paragraph{BH} An extended form of the Buchm\"uller-Haidt parametrization~\cite{ref:BH}:
$$f(l)=A+\ln\left(\frac{Q^2}{Q^2_0}+P_1\right)\ln\left(\frac{x_0}{x}\right) $$
giving the 6 free parameters: $\sigma_0,M^2,A,Q^2_0,P_1,x_0$.  Note that Buchm\"uller and Haidt fit the $F_2$ data directly, and only considered the range $Q^2>5$~GeV$^2$.  The parametrization used here is extended in that the parameters $M^2, P_1$ are used to give a smooth transition to the photoproduction region.  The best fit parameters found in ~\cite{ref:BH} would indicate that
\begin{itemize}
\item[$A$] should be near the ratio $a/m=0.078/0.364\approx 0.2$ of the parameters from~\cite{ref:BH}, although the value will be strongly correlated to the other fit parameters;
\item[$Q^2_0, P_1$]  should be of order $1$;
\item[$x_0$] was found to be $0.074$ in \cite{ref:BH}, and should be near this value in this parametrization also.
\end{itemize}

This is in no way an exhaustive list of parametrizations which have been tried out for the small-$x$ data, but the parametrizations given here are intended to be a representative set to guide the discussion on the features of the data.
Many other parametrizations have also been used, and found to give good results, so a good fit from one of the parametrizations discussed above does not guarantee that it contains the right physics motivation.

\section{Fitting Technique}
The parameter value extraction is based on the learning rule:

$$P(\vec{\lambda}|D,I_0) \propto P(D|\vec{\lambda},I_0) P_0(\vec{\lambda}|I_0) $$
where the term on the left is the posterior probability density for the fit parameters, $\vec{\lambda}$, given the data, $D$, and all assumptions, $I_0$, (including model chosen).  The terms on the right are the probability density of the data given the model and parameters (likelihood) and the prior probability density for the parameters given the model chosen.   The posterior is normalized at the end, yielding the famous result associated with Bayes theorem:
\begin{equation}
\label{eq:Bayes}
P(\vec{\lambda}|D,I_0) =\frac{P(D|\vec{\lambda},I_0) P_0(\vec{\lambda}|I_0)}{\int P(D|\vec{\lambda},I_0) P_0(\vec{\lambda}|I_0) d\vec{\lambda}} 
\end{equation}

\begin{figure}[hbpt]
\begin{center}
   \includegraphics[width=0.8\textwidth]{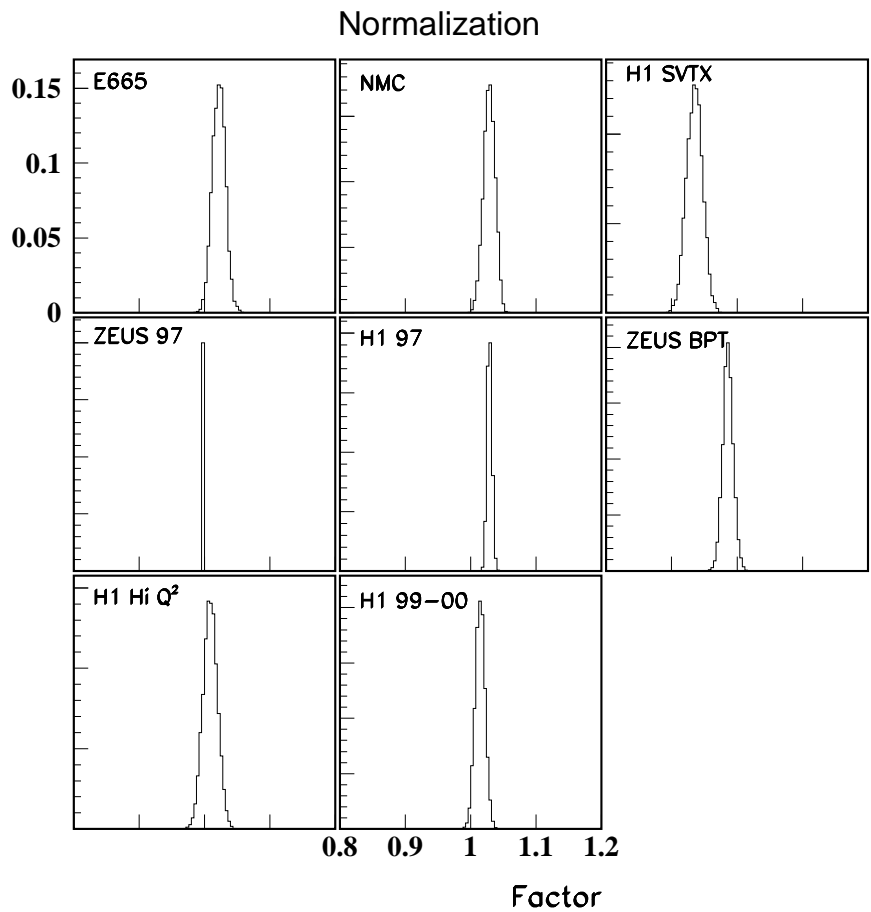}
\caption{\it Probability density distribution for the normalizations of the different data sets used in the fitting.  These normalizations are for fit 22.}
\label{fig:norms}
\end{center}
\end{figure}

\begin{figure}[hbpt]
\begin{center}
   \includegraphics[width=0.6\textwidth]{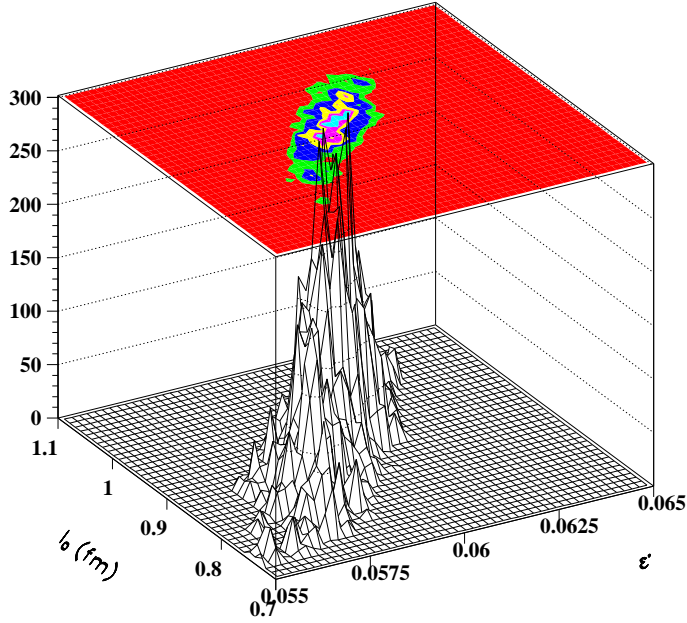}
\caption{\it Probability density distribution and contours for $P(l_0,\epsilon'|D,I_0)$ for fit 22.}
\label{fig:lalphap}
\end{center}
\end{figure}

The prior probabilities are chosen to be flat for all model parameters.  As described above, the normalizations for each data set, except ZEUS97, are left free and the priors are Gaussians with widths specified in section~\ref{sec:data}.   The likelihood of the data is evaluated assuming a Gaussian probability distribution for fluctuations from the model expectation, with the Gaussian variance given either by Eq.~\ref{eq:uncorr} or Eq.~\ref{eq:quad}.  For each data point, the model prediction, $F_2^{pred}$, is compared to the data point, $F_{2,i}$, scaled by the normalization constant, $S_{\rm expt}$, and, in some fits, the $y$ dependent factor to account for correlated uncertainties. The log of the probability density for one data point is calculated as
$$\ln{P(F_{2,i}|\vec{\lambda},I_0)}=\ln{\frac{1}{\sqrt{2\pi e_i^2}}}-0.5\left( \frac{F_2^{pred}-F_{2,i} f(y)  S_{\rm expt}}{e_i^2}\right)^2 \;\; .
$$
The log of the total probability density for the data is given by the sum of the contributions from the individual data points. This is then added to the log of the probability density for the prior to give the (unnormalized) value of $\ln{P(\vec{\lambda}|D,I_0)}$.

The total number of fit parameters ranged from a minimum of $13$ for the model parameters and free normalizations, to $27$, if all parameters related to the data sets were also allowed to vary.  The parameters related to the data sets are so-called nuisance parameters, and can be integrated out if needed:
\begin{equation}
\label{eq:params}
P(\vec{\lambda}_{\rm model}|D,I_0) = \int P(\vec{\lambda}|D,I_0) d\vec{\lambda}_{\rm data}
\end{equation}
where $\vec{\lambda}=\{ \vec{\lambda}_{\rm model},\vec{\lambda}_{\rm data}\}$
\subsection{Markov Chain Monte Carlo}
The posterior probability, Eq.~\ref{eq:Bayes}, is determined using a Markov Chain Monte Carlo (MCMC)~\cite{ref:Markov}.  
For each parameter, the allowed range, starting value, and proposal function are set.  The proposal function for varying the parameters was always chosen flat.  The MCMC implementation automatically changed the sampling range of the proposal function after an initial number of iterations, in order to have the efficiency for keeping the proposed parameter value in the range 10-50\%.  The MCMC was run until the mean and rms of $P(\vec{\lambda}|D,I_0) $ over sufficiently long periods was stable.  Once convergence was achieved, the MCMC was run to save the distribution of $P(\vec{\lambda}|D,I_0)$ (individual instances of $\vec{\lambda}$ were saved). This gave access to the full probability density distribution. Any function of the parameters can then be evaluated, as well as the probability density for this function, without approximations.  The integral in Eq.~\ref{eq:Bayes} is not solved for explicitely.  Rather, the MCMC guarantees sampling of $P(\vec{\lambda}|D,I_0)$ according to the correct probability density, so that the normalization is given by the total number of events in the chain (after convergence).

As an example, the normalizations for the different data sets for fit 22 (see Table~\ref{tab:fits}) are shown in Fig.~\ref{fig:norms}.   The results for the normalizations from the different fits are given in the next section. 

The correlations amongst the fit parameters are available via the posterior probability density.  As an example, the probability density contours for the marginalized density
$$P(l_0,\epsilon'|D,I_0)=\int P(\sigma_0,M^2,l_0,\epsilon_0,\epsilon',Q^2_0,\vec{\lambda}_{\rm data}|D,I_0) d\sigma_0 dM^2 d\epsilon_0 dQ^2_0 d\vec{\lambda}_{\rm data} $$
for the fit 22 are shown in Fig.~\ref{fig:lalphap}.

\subsection{Goodness-of-fit}
The goodness-of-fit was evaluated using a `p-value' defined as follows:
\begin{equation}
\label{eq:pvalue}
p=\frac{\int_{P^*(\vec{x})<P^D} P^*(\vec{x}) d\vec{\lambda}_{\rm data}d\vec{x}}{\int P^*(\vec{x}) d\vec{\lambda}_{\rm data}d\vec{x}}
\end{equation}
where
\begin{eqnarray*}
P^*(\vec{x}) & = &P(\vec{x}|\vec{\lambda}^*_{\rm model},\vec{\lambda}_{\rm data},I_0)P_0(\vec{\lambda}_{\rm data}|I_0) \\
P^D & = &P(D|\vec{\lambda}^*_{\rm model},\vec{\lambda}^*_{\rm data},I_0)P_0(\vec{\lambda}^*_{\rm data}|I_0) \;\; .
\end{eqnarray*}
Here $\vec{x}$ is a possible realization of the data, $D$ is the observed data, and $\vec{\lambda}^*_{\rm model},\vec{\lambda}^*_{\rm data}$ represent the values of the model and data parameters which maximize  $P(\vec{\lambda}|D,I_0)$.  Note that only the model parameters are fixed at the values giving the mode for $P^*(\vec{x})$, while the nuisance parameters are allowed to vary according to their priors. This p-value is a tail area probability, and is just the probability that an experiment would observe a smaller (likelihood$\cdot$data prior) than the one found, assuming the model is correct.  If the likelihood accurately describes nature and the experimental effects, then the p-value distribution for many experiments would be flat between $(0,1)$. Small values therefore suggest that the model is unlikely to be a good representation of nature (assuming the modelling of the experimental conditions was done correctly), since this says that few experiments are expected to have such small values of $P^D$.  In practice, the p-value was calculated using 10000 simulated experiments where the normalizations were allowed to vary according to the probability density in the priors, and the data points were fluctuated around the expectation value with a probability density given by a Gaussian with appropriate width.  The fraction of simulated experiments satisfying the condition given on the integral in Eq.~\ref{eq:pvalue} was then calculated. 

\subsection{Bin-centering and further analysis}
\label{sec:center}
The data from the different experiments are reported in some cases at fixed values of $x$ and varying $Q^2$, in other cases at fixed $Q^2$ and varying $x$, or at fixed $Q^2$ and varying $y$.  This is no problem for the model fitting procedure, but it does make data presentation cumbersome.  The data were therefore `bin-centered' by moving data to fixed values of $Q^2$ using the parametrization under study, as follows:
\begin{equation}
\label{eq:bincenter}
F_2(Q^2_c,x) = \frac{F_2^{\rm pred}(Q^2_c,x)}{F_2^{\rm pred}(Q^2,x)}F_2(Q^2,x)f(y)S_{\rm expt} \;.
\end{equation}
where $f(y)$ is defined in Eq.~\ref{eq:systs} and $S_{\rm expt} $ is the normalization factor for the experiment in question.  The following $Q^2_c$ values have been used: 0.15, 0.4, 0.8, 1.5, 3, 5, 7.5, 10, 15, 20, 30, 40, 50, 70, 100, 150, 200, 250, 300, 400, 500, 600, 750, 1000~GeV$^2$.

The MCMC was then rerun for the bin-centered data at each $Q^2_c$ value using the simple parametrization:
\begin{equation}
\label{eq:refit}
\sigma^{\gamma P}(l)=\sigma^{\gamma P}_0 l^{\lambda_{\rm eff}}
\end{equation}

\noindent and the probability density $P(\sigma^{\gamma P}_0 ,\lambda_{\rm eff}|D,I_0)$ was evaluated separately in each $Q^2_c$ bin.

\begin{figure}[hbpt]
\begin{center}
   \includegraphics[width=1.0\textwidth]{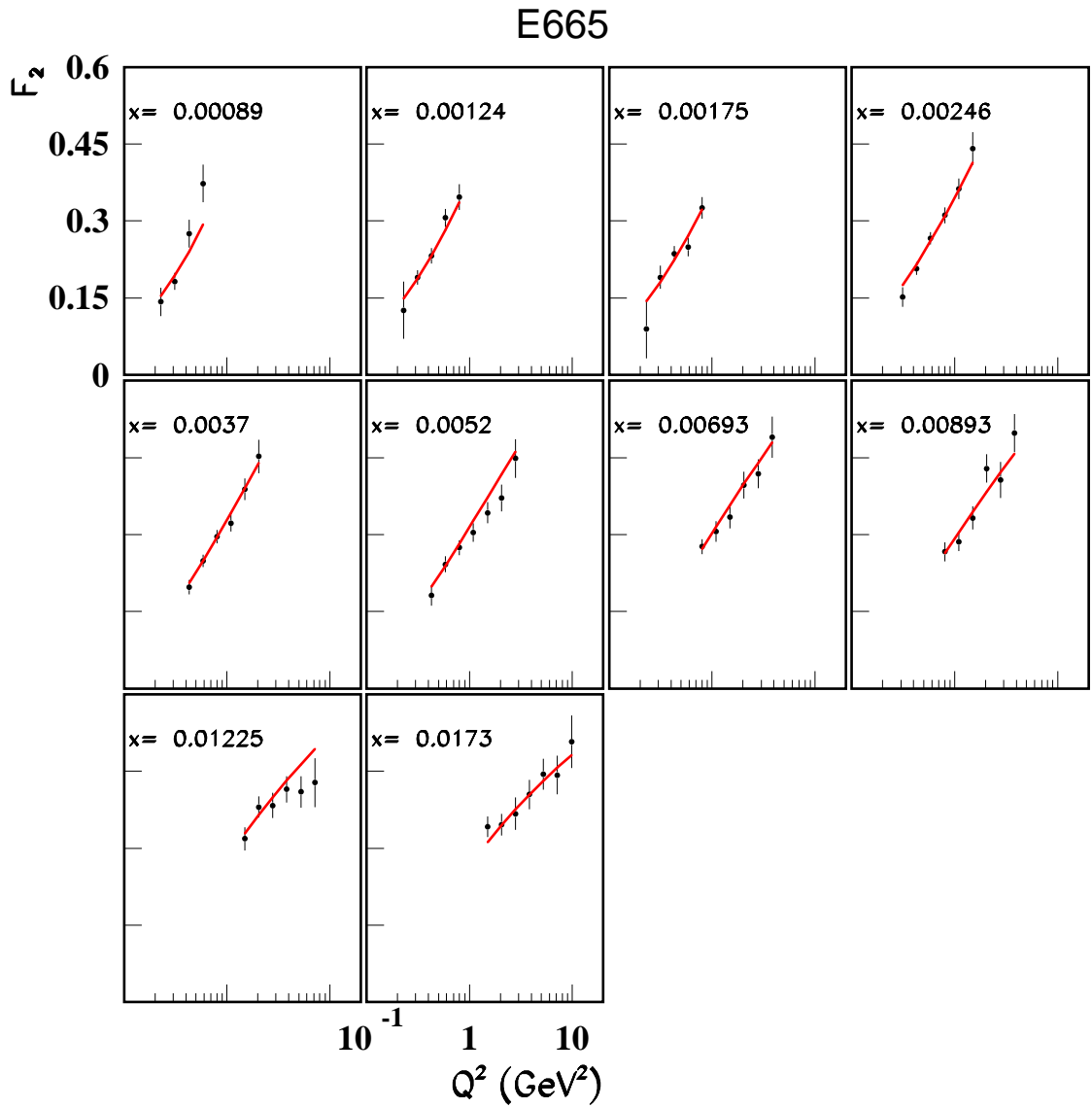}
\caption{\it The  E665 $F_2$ data with the D parametrization superposed (fit 22) .}
\label{fig:e665}
\end{center}
\end{figure}

\begin{figure}[hbpt]
\begin{center}
   \includegraphics[width=1.0\textwidth]{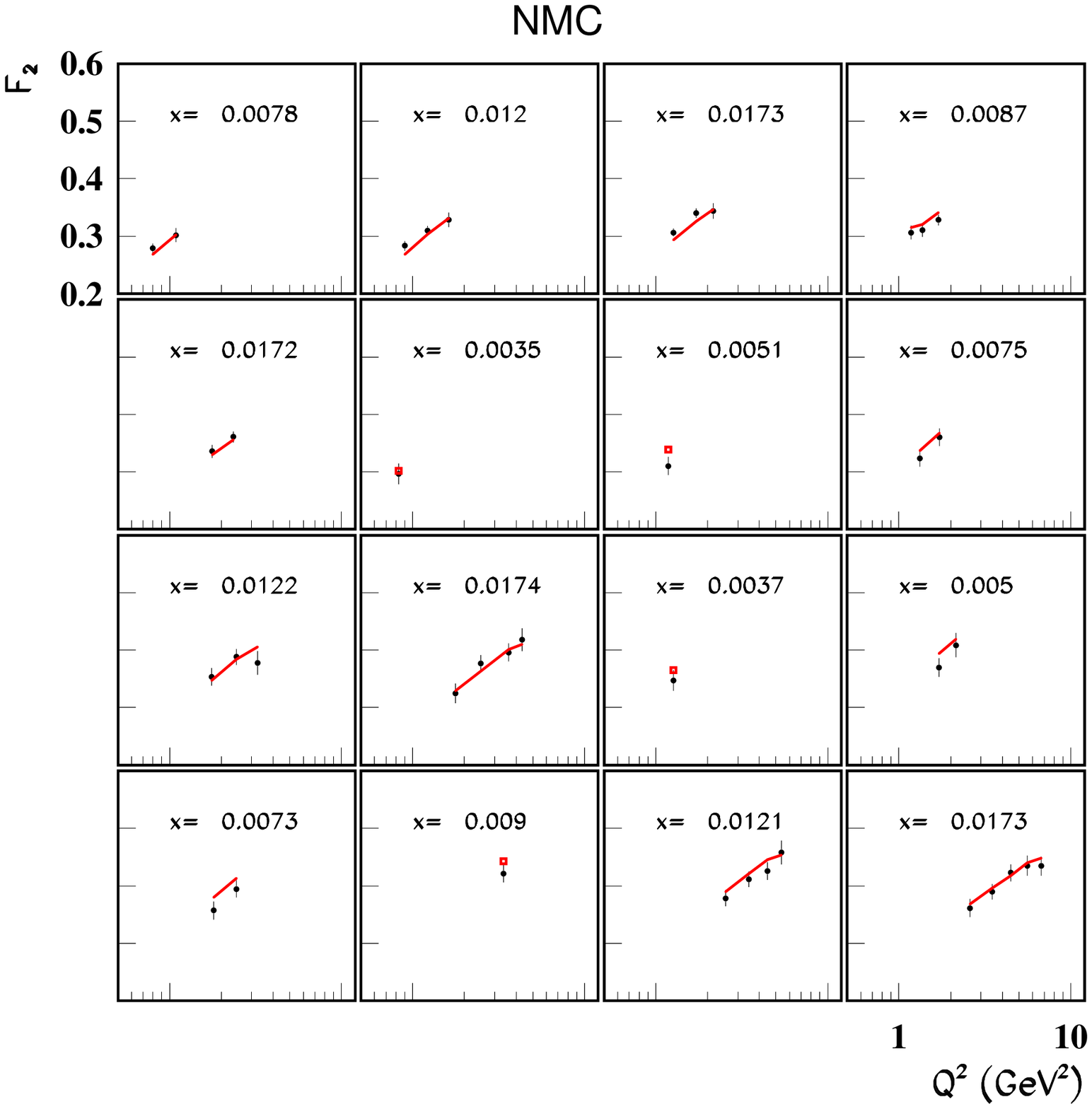}
\caption{\it The  NMC $F_2$ data with the D parametrization superposed (fit 22). 
In bins where only one data point is available, the prediction is shown with a square.}
\label{fig:nmc}
\end{center}
\end{figure}

\begin{figure}[hbpt]
\begin{center}
   \includegraphics[width=1.0\textwidth]{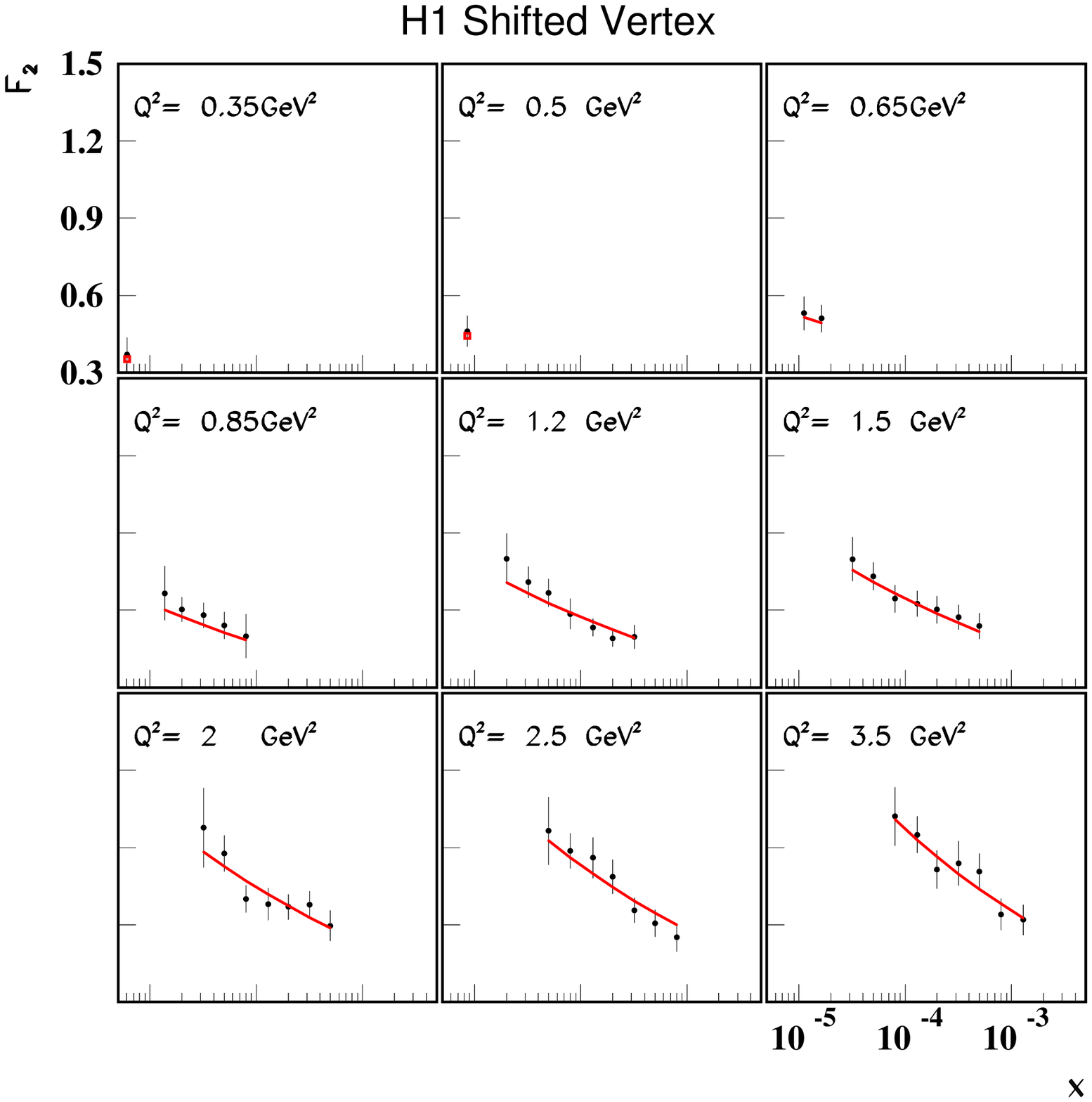}
\caption{\it The  H1 SVTX $F_2$ data with the D parametrization superposed (fit 22).
In bins where only one data point is available, the prediction is shown with a square.}
\label{fig:h1svtx}
\end{center}
\end{figure}

\begin{figure}[hbpt]
\begin{center}
   \includegraphics[width=1.0\textwidth]{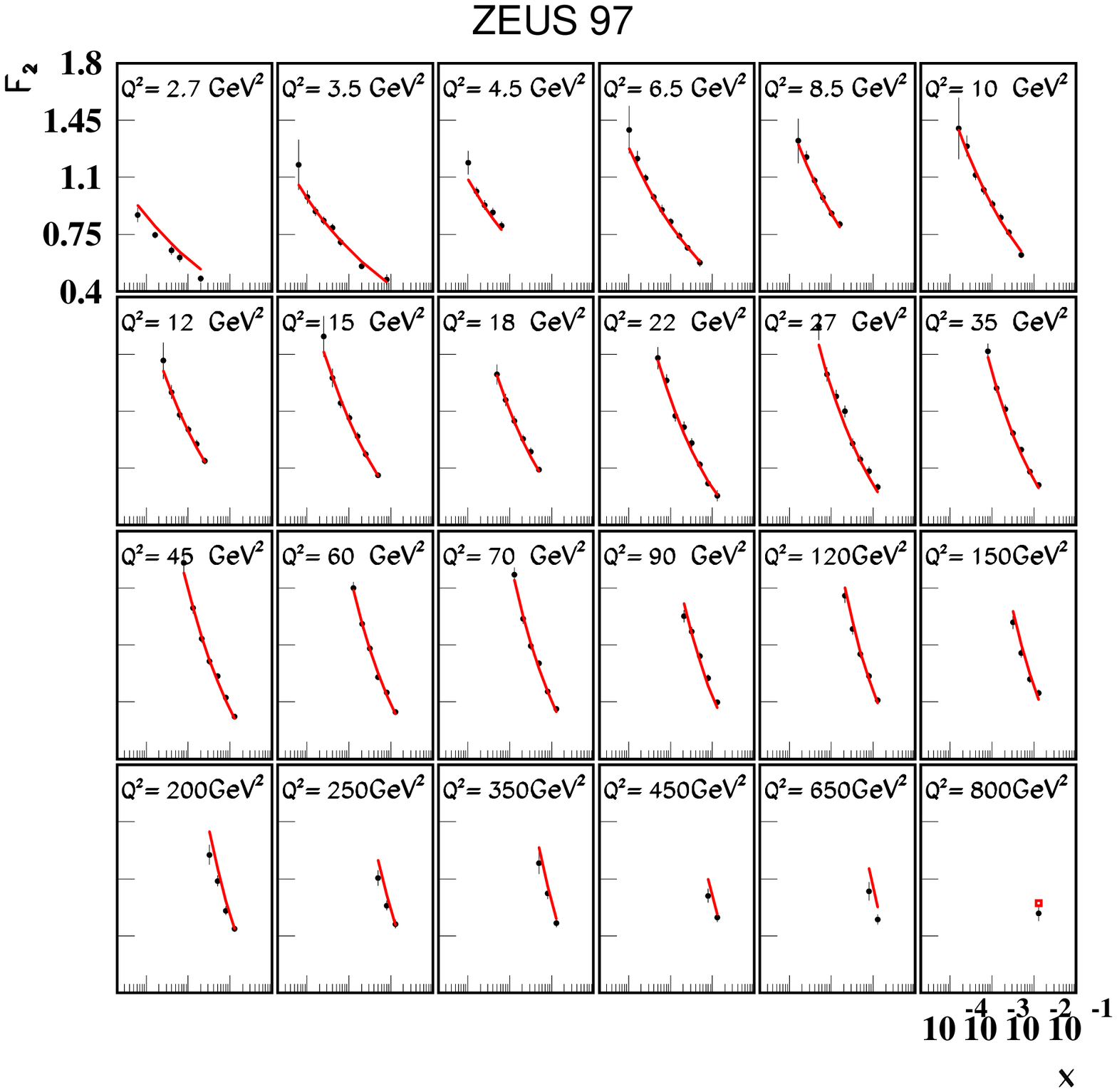}
\caption{\it The  ZEUS 97 $F_2$ data with the D parametrization superposed (fit 22).
In bins where only one data point is available, the prediction is shown with a square.}
\label{fig:zeus97}
\end{center}
\end{figure}

\begin{figure}[hbpt]
\begin{center}
   \includegraphics[width=1.0\textwidth]{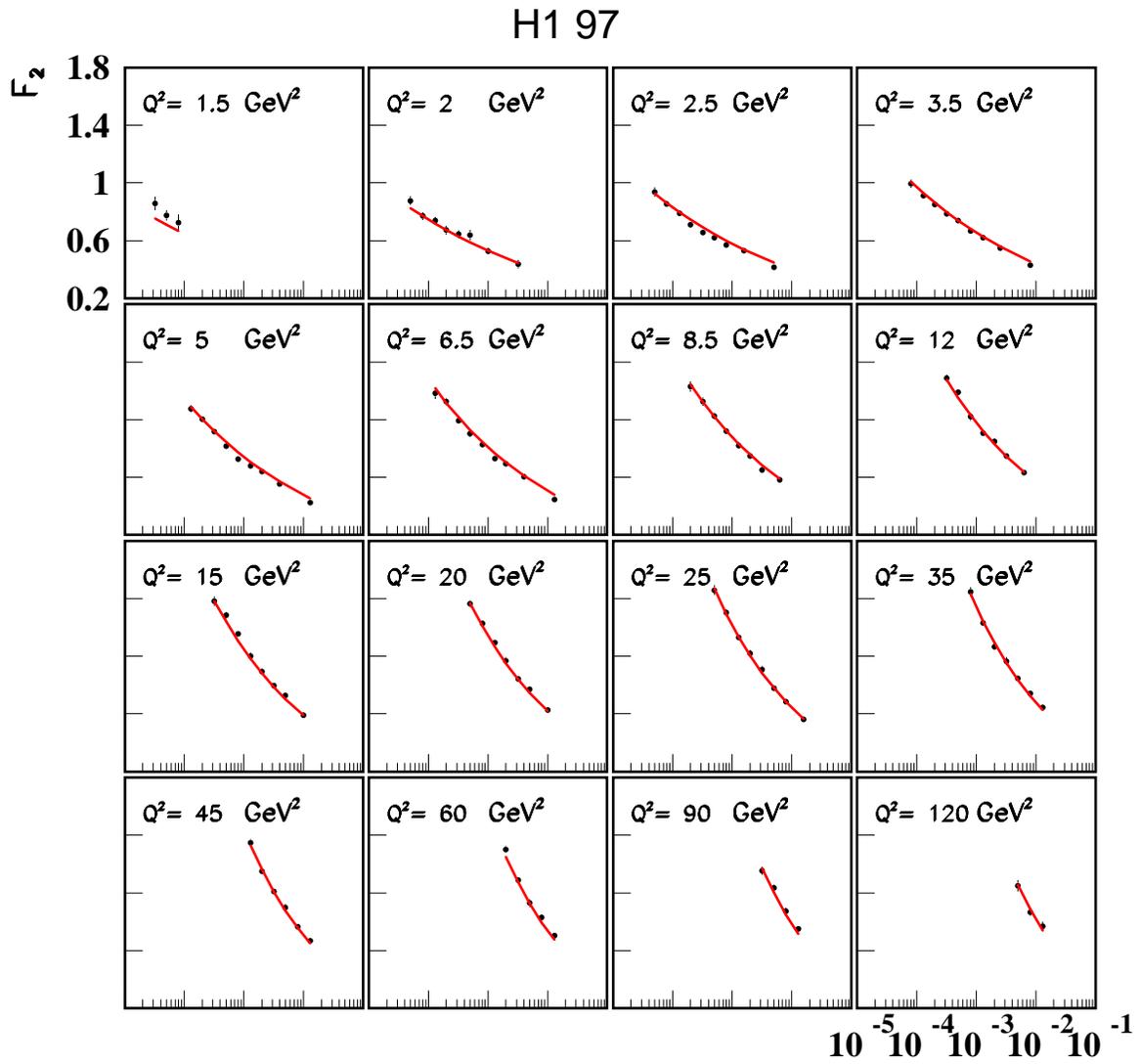}
\caption{\it The  H1 97 $F_2$ data with the D parametrization superposed (fit 22).}
\label{fig:h197}
\end{center}
\end{figure}

\begin{figure}[hbpt]
\begin{center}
   \includegraphics[width=1.0\textwidth]{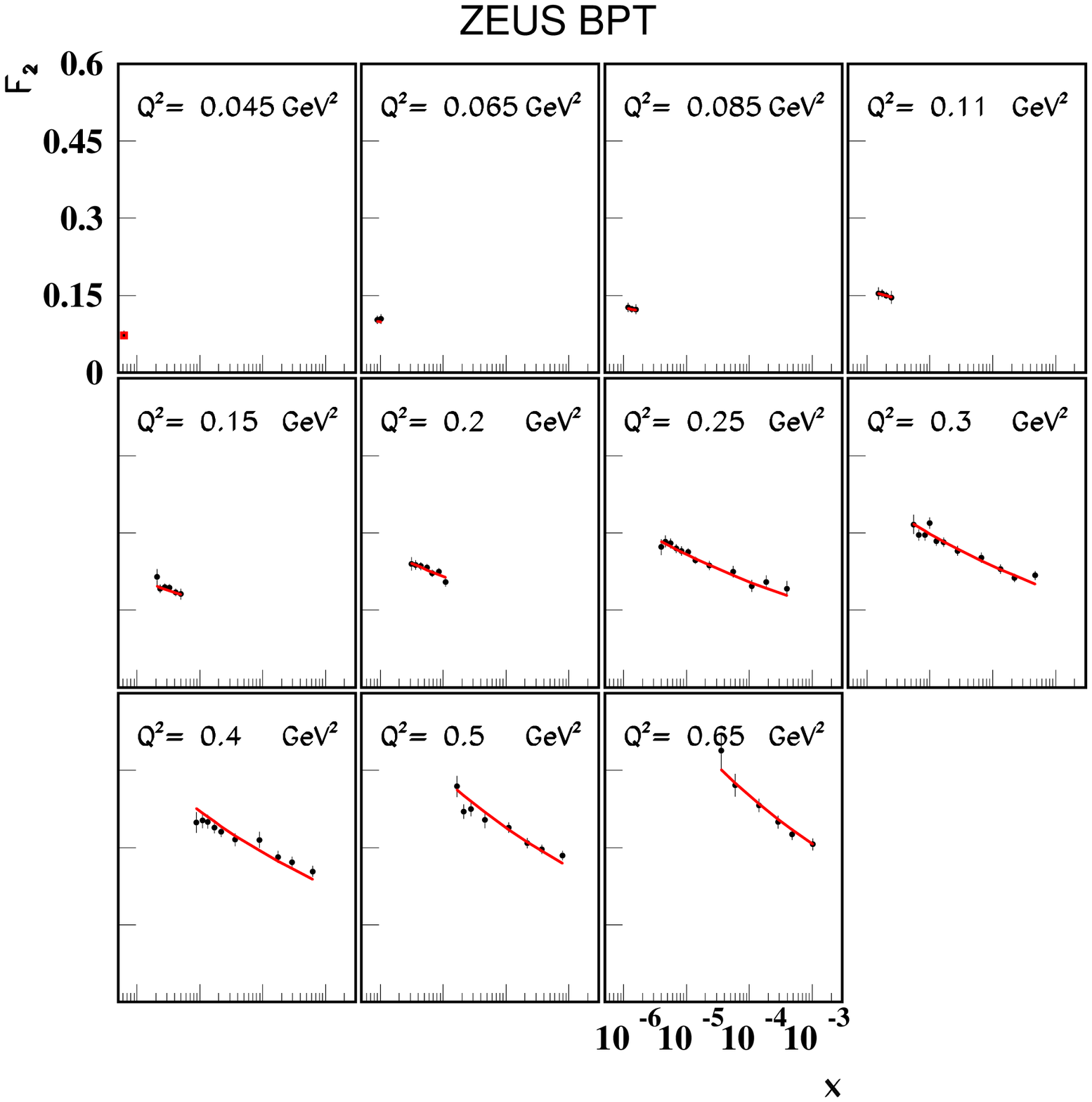}
\caption{\it The  ZEUS BPT $F_2$ data with the D parametrization superposed (fit 22).
In bins where only one data point is available, the prediction is shown with a square.}
\label{fig:zeusbpt}
\end{center}
\end{figure}

\begin{figure}[hbpt]
\begin{center}
   \includegraphics[width=1.0\textwidth]{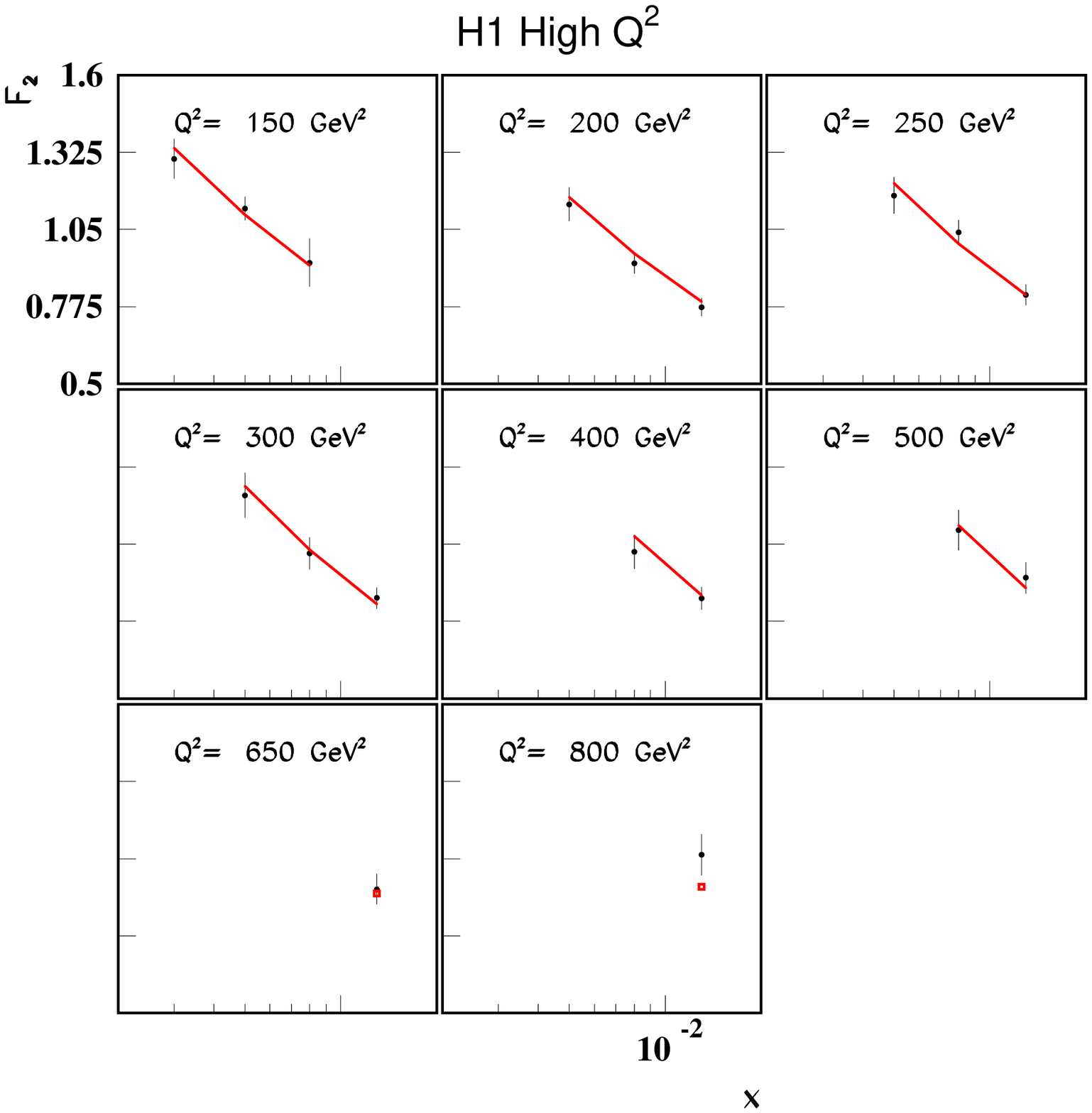}
\caption{\it The  H1 hiQ $F_2$ data with the D parametrization superposed (fit 22).
In bins where only one data point is available, the prediction is shown with a square.}
\label{fig:h1hiq}
\end{center}
\end{figure}

\begin{figure}[hbpt]
\begin{center}
   \includegraphics[width=1.0\textwidth]{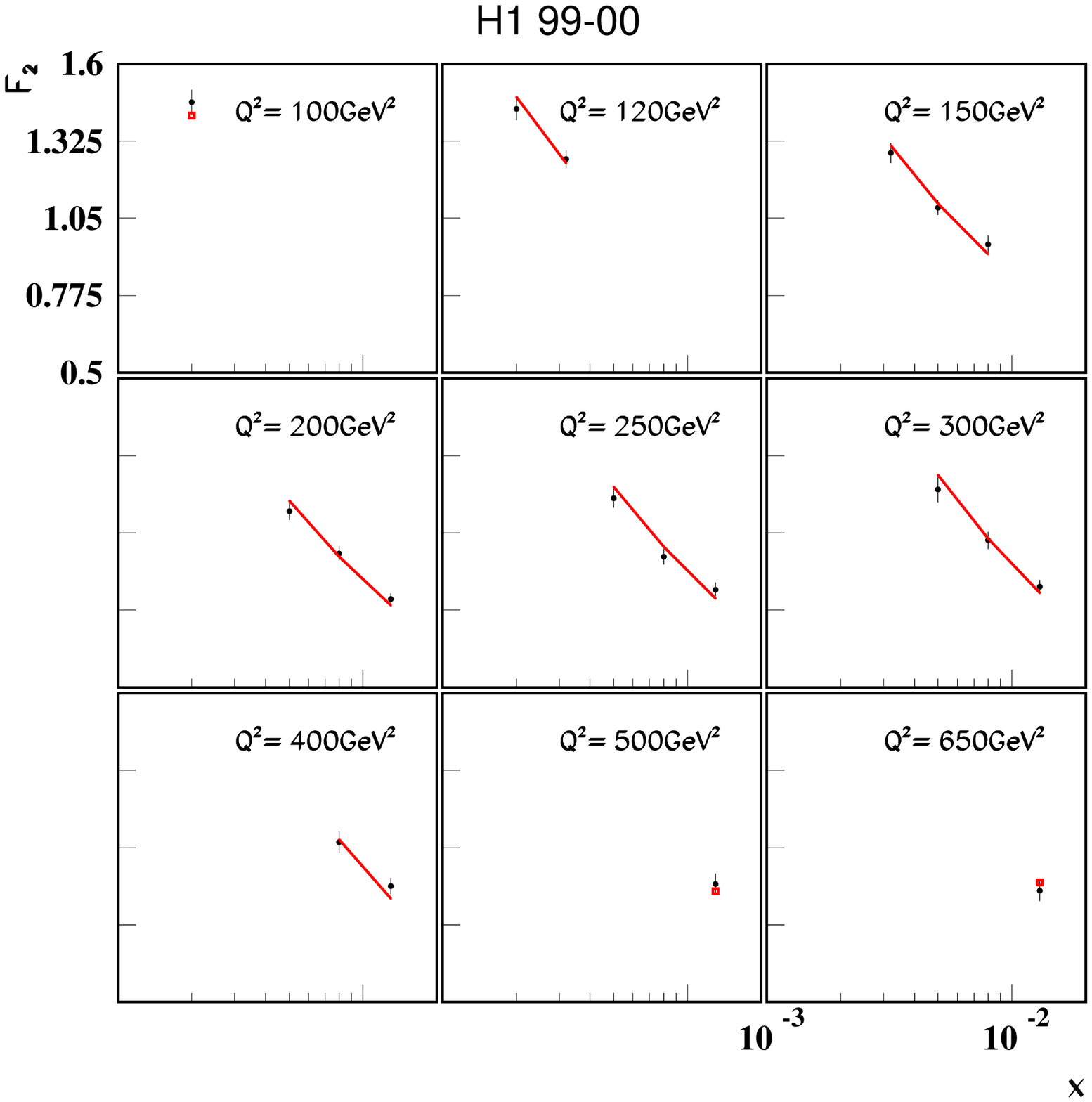}
\caption{\it The  H1 9900 $F_2$ data with the D parametrization superposed (fit 22).
In bins where only one data point is available, the prediction is shown with a square.}
\label{fig:h19900}
\end{center}
\end{figure}

\section{Results}
Fits were performed using the parametrizations described in Section~\ref{sec:params}. The kinematic range over which the fits were performed was also varied.  The results were stable for $x<0.025$, but clearly deteriorated if the upper limit for $x$ was increased to $x<0.05$.  The best fits were achieved with the D and 2P parametrizations, while the BH parametrization was less successful at describing the data.  

\begin{table}[htdp]
\caption{\it Summary of the fit results for different parametrizations.  The first column gives the fit number, which is used as a label in subsequent tables and plots. In cases where no value of $x_{min}$ is specified, all data with $x<x_{max}$ were used.  If no value of $y_{max}$ is given, then data up to the largest available $y$ were used. The p-value, indicating the goodness-of-fit, is given in the last column.  There is no p-value for fit 26 since in this case no global fit was performed (only the extraction of $\lambda_{\rm eff}$).}
\begin{center}
\begin{tabular}{|c|c|cccc|}
\hline
Fit & Parametrization & $x_{min}$ & $x_{max}$ & $y_{max}$ & $p$ \\
\hline
27 &D & & $0.01$   &   & $0.89$ \\
22 &       & & $0.02$ & & $0.79$ \\
24 &       &  & $0.025$   & & $0.22 $ \\
15 &    ($Q^2_0=M^2$)     &  & $0.02$ & &  $0.08$ \\
31 & & & $0.01$  &  $0.3$ & $0.95$ \\
\hline 
28 & 2P & & $0.01$ & &  $1.0$ \\
21 &       & & $0.02$  & &  $0.99$ \\
25 &       &   & $0.025$  & &  $0.91$ \\
26 &       & $0.002$  & $0.025$ &  &  \\
\hline 
29 &  BH  & & $0.02$ & & $0.03$ \\
\hline 
 \end{tabular}
\end{center}
\label{tab:fits}
\end{table}%

A summary of the fit results is given in Table~\ref{tab:fits}, while the fitted parameter values for some reference fits are given in Table~\ref{tab:Dpars}-\ref{tab:BHpars}.  Each fit is assigned a number which will be used in the remainder of the document as a reference number. 

The p-values clearly distinguish between the D and 2P models on the one hand, and the BH and other models (not shown) which have been fitted to the data.  The p-values are very close to 1 for the D and 2P fits, indicating that adding the statistical and systematic errors in quadrature as done in Eq.~\ref{eq:quad} overestimates the uncertainties.  On the other hand, handling the y-dependent systematic effects as in Eq.~\ref{eq:uncorr} resulted in small probabilities for the fit results (typically 0).  However, the fitted parameter values changed only slightly.  No results are quoted for the fits where an attempt has been made to subtract the correlated systematics uncertainties.  The fitted data are shown in Figs~\ref{fig:e665}-~\ref{fig:h19900} together with the D parametrization (fit 22) for the case $x<0.02$ to give a sense of the quality of the fits.  

Several quantities are calculated from the Markov Chain output and reported in the tables. The definitions are as follows:
\paragraph{mode of $\lambda_i$} The value of $\lambda_i$ which maximizes the marginalized posterior probability density
$$\stackrel{\lambda_i}{\max}\{ P(\lambda_i|D,I_0) = \int P(\vec{\lambda}|D,I_0) d\vec{\lambda}_{j\neq i} \}$$
\paragraph{Mean of $\lambda_i$} The expectation value
$$<\lambda_i> = \int P(\lambda_i|D,I_0) \lambda_i d\lambda_{i}$$
\paragraph{Median of $\lambda_i$} The value of $\lambda_i$ such that 50~\% of the probabilty is below this value
$$ \int_{\lambda_{min}}^{\lambda_{med}} P(\lambda_i|D,I_0) d\lambda_{i}=0.5$$
where $\lambda_{min}$ is the minimum possible value for parameter $\lambda_i$.  The desired value is $\lambda_{med}$.
\paragraph{Central Interval} The $(1-2\alpha)$ central interval is defined such that a fraction $\alpha$ of the probability is contained on either side of the interval
$$ \alpha=\int_{\lambda_{min}}^{\lambda_{lower}} P(\lambda_i|D,I_0) d\lambda_{i}=
\int_{\lambda_{upper}}^{\lambda_{max}} P(\lambda_i|D,I_0) d\lambda_{i}$$
where the desired interval is $[\lambda_{lower},\lambda_{upper}]$.  The minimum and maximum values of the parameter are $\lambda_{min}, \lambda_{max}$.
\paragraph{rms} The root-mean-square is defined as usual
$$rms = \sqrt{ \left[ \int P(\lambda_i|D,I_0) \lambda_i^2 d\lambda_{i} - \left(\int P(\lambda_i|D,I_0) \lambda_i d\lambda_{i}\right)^2 \right]} $$

\begin{table}[htdp]
\caption{\it Parameter results for the D parametrization for fit 27.}
\begin{center}
\begin{tabular}{|c|c|ccccc|}
\hline
Fit & Parameter & Mode & Mean & Median & 68~\% central range & rms \\
\hline
27   &  $\sigma_0$~(mbarn) & $0.0627$ & $0.0641$ & $0.0635$ & $0.0616-0.0669$ & $0.0026$ \\
27   &  $M^2$~(GeV$^2$) & $0.635$ & $0.630$ & $0.632$ & $0.591-0.665$ & $0.034$ \\
27   &  $l_0$~(fm) & $1.22$ & $1.19$ & $1.19$ & $1.06-1.30$ & $0.118$ \\
27   &  $\epsilon_0$ & $0.0671$ & $0.0674$ & $0.0671$ & $0.0613-0.0737$ & $0.0062$ \\
27   &  $\epsilon'$ & $0.0636$ & $0.0637$ & $0.0626$ & $0.0606-0.0646$ & $0.0020$ \\
27   &  $Q^2_0$~GeV$^2$ & $1.54$ & $1.44$ & $1.39$ & $1.19-1.54$ & $0.17$ \\
\hline
27   &  E665 norm & $1.017$ & $1.016$ & $1.015$ & $1.004-1.027$ & $0.011$ \\
27   & NMC norm & $1.042$ & $1.044$ & $1.044$ &  $1.031-1.056$ & $0.014$ \\
27   & H1 SVTX norm & $0.933$ & $0.934$ & $0.933$ &  $0.921-0.945$ & $0.012$ \\
27   & H197 norm & $1.030$ & $1.030$ & $1.030$ &  $1.026-1.033$ & $0.004$ \\
27   & ZEUS BPT norm & $0.985$ & $0.983$ & $0.983$ &  $0.974-0.991$ & $0.008$ \\
27   & H1 hiQ norm & $1.009$ & $1.007$ & $1.007$ &  $0.994-1.019$ & $0.013$ \\
27   & H1 9900 norm & $1.015$ & $1.017$ & $1.017$ &  $1.008-1.025$ & $0.008$ \\
\hline
 \end{tabular}
\end{center}
\label{tab:Dpars}
\end{table}%

\begin{table}[htdp]
\caption{\it Parameter results for the 2P parametrization for fit 28.}
\begin{center}
\begin{tabular}{|c|c|ccccc|}
\hline
Fit & Parameter & Mode & Mean & Median & 68~\% central range & rms \\
\hline
28   &  $\sigma_0$~(mbarn) & $0.0583$ & $0.0579$ & $0.0577$ & $0.0555-0.0601$ & $0.0024$ \\
28   &  $M^2$~(GeV$^2$) & $0.581$ & $0.576$ & $0.579$ & $0.539-0.605$ & $0.032$ \\
28   &  $l_0$~(fm) & $0.538$ & $0.505$ & $0.507$ & $0.428-0.577$ & $0.075$ \\
28   &  $\epsilon_0$ & $0.0855$ & $0.0836$ & $0.0837$ & $0.080-0.087$ & $0.0033$ \\
28   &  $\epsilon_1$ & $0.370$ & $0.366$ & $0.366$ & $0.358-0.373$ & $0.008$ \\
28   &  $\Lambda^2$~GeV$^2$  & $31.6$ & $31.6$ & $31.5$ & $28.8-34.3$ & $2.7$ \\
\hline
28   &  E665 norm & $1.000$ & $1.004$ & $1.004$ & $0.993-1.015$ & $0.011$ \\
28   & NMC norm & $1.027$ & $1.025$ & $1.025$ &  $1.012-1.037$ & $0.013$ \\
28   & H1 SVTX norm & $0.936$ & $0.935$ & $0.935$ &  $0.923-0.947$ & $0.012$ \\
28   & H1 97 norm & $1.033$ & $1.035$ & $1.034$ &  $1.031-1.038$ & $0.004$ \\
28   & ZEUS BPT norm & $1.000$ & $1.000$ & $1.000$ &  $0.991-1.009$ & $0.009$ \\
28   & H1 hiQ norm & $0.965$ & $0.964$ & $0.964$ &  $0.951-0.977$ & $0.013$ \\
28   & H1 9900 norm & $0.970$ & $0.968$ & $0.967$ &  $0.958-0.976$ & $0.009$ \\
\hline
 \end{tabular}
\end{center}
\label{tab:2Ppars}
\end{table}%

\begin{table}[htdp]
\caption{\it Parameter results for the BH parametrization for fit 29.}
\begin{center}
\begin{tabular}{|c|c|ccccc|}
\hline
Fit & Parameter & Mode & Mean & Median & 68~\% central range & rms \\
\hline
29   &  $\sigma_0$~(mbarn) & $0.0320$ & $0.0330$ & $0.0326$ & $0.0298-0.0367$ & $0.0032$ \\
29   &  $M^2$~(GeV$^2$) & $0.273$ & $0.275$ & $0.271$ & $0.241-0.305$ & $0.029$ \\
29   &  $Q^2_0$~(GeV$^2$) & $0.930$ & $0.938$ & $0.934$ & $0.871-1.001$ & $0.065$ \\
29   &  $P1$ & $1.35$ & $1.38$ & $1.37$ & $1.31-1.44$ & $0.06$ \\
29   &  $x_0$ & $0.0442$ & $0.0448$ & $0.0446$ & $0.0421-0.0470$ & $0.0025$ \\
29   &  $A$ & $2.69$ & $2.68$ & $2.67$ & $2.58-2.77$ & $0.10$ \\
\hline
29   &  E665 norm & $0.975$ & $0.975$ & $0.975$ & $0.964-0.987$ & $0.012$ \\
29   & NMC norm & $0.953$ & $0.953$ & $0.953$ &  $0.940-0.967$ & $0.014$ \\
29   & H1 SVTX norm & $0.921$ & $0.921$ & $0.921$ &  $0.909-0.933$ & $0.013$ \\
29   & H1 97 norm & $1.022$ & $1.027$ & $1.027$ &  $1.023-1.030$ & $0.004$ \\
29   & ZEUS BPT norm & $0.997$ & $0.995$ & $0.995$ &  $0.987-1.004$ & $0.008$ \\
29   & H1 hiQ norm & $0.998$ & $1.000$ & $1.000$ &  $0.988-1.012$ & $0.013$ \\
29   & H1 9900 norm & $1.006$ & $1.005$ & $1.005$ &  $0.997-1.013$ & $0.008$ \\
\hline
 \end{tabular}
\end{center}
\label{tab:BHpars}
\end{table}%

The parameter estimates were found to be near the expected values, and no parameter was at the limit of the allowed range.  The parameter values for a given parametrization did not show strong variations when changing the kinematic range for the fits.  Larger variations were seen from changing the parametrization from one form to another.  As is clearly seen in the tables, the fitted normalization values for the different data sets depends strongly on the parametrization chosen, particularly for the data at the lowest and highest $Q^2$.  These variations are larger than the fit uncertainty on the normalization in individual fits.  The only data set which consistently requires a significant normalization correction is the H1 SVTX data, where a normalization factor of typically $0.93$ is preferred by the fits.

The mode, mean and median of the fit values are very close, indicating that the posterior probability distributions for these parameters is symmetric.  Also, the 68~\% central range is very nearly twice the rms value, indicating that the distributions are Gaussian in shape.

\begin{figure}[hbpt]
\begin{center}
   \includegraphics[width=1.\textwidth]{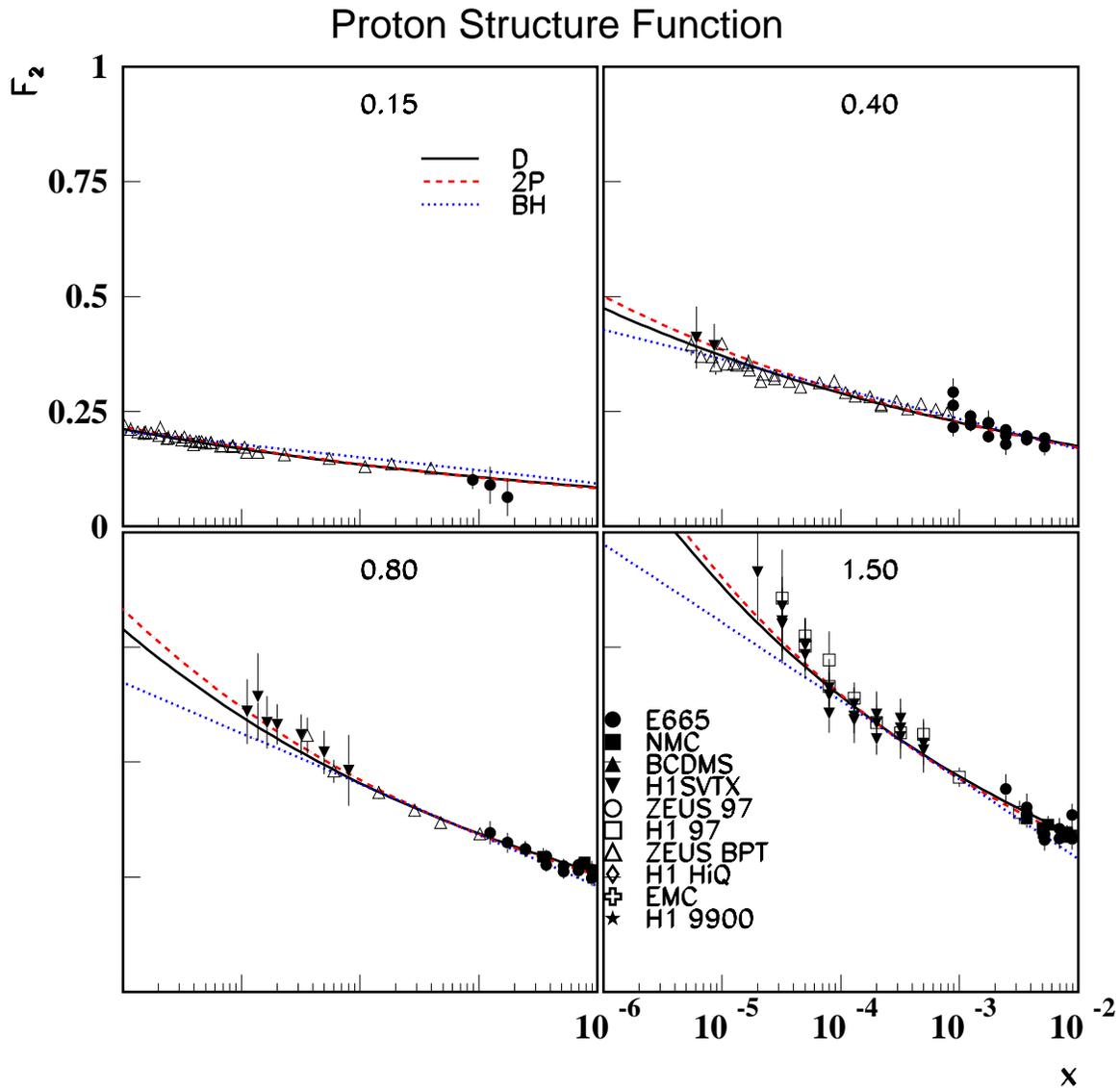}
\caption{\it Comparison of parametrizations to bin-centered $F_2$ data for $x<0.01$.  The values near the top of each box give the $Q^2$ value in GeV$^2$ to which the data have been moved.  The symbols for the different data sets are given in the lower right plot, while the different parametrizations are defined in the upper left plot.}
\label{fig:f2center1}
\end{center}
\end{figure}

\begin{figure}[hbpt]
\begin{center}
   \includegraphics[width=1.\textwidth]{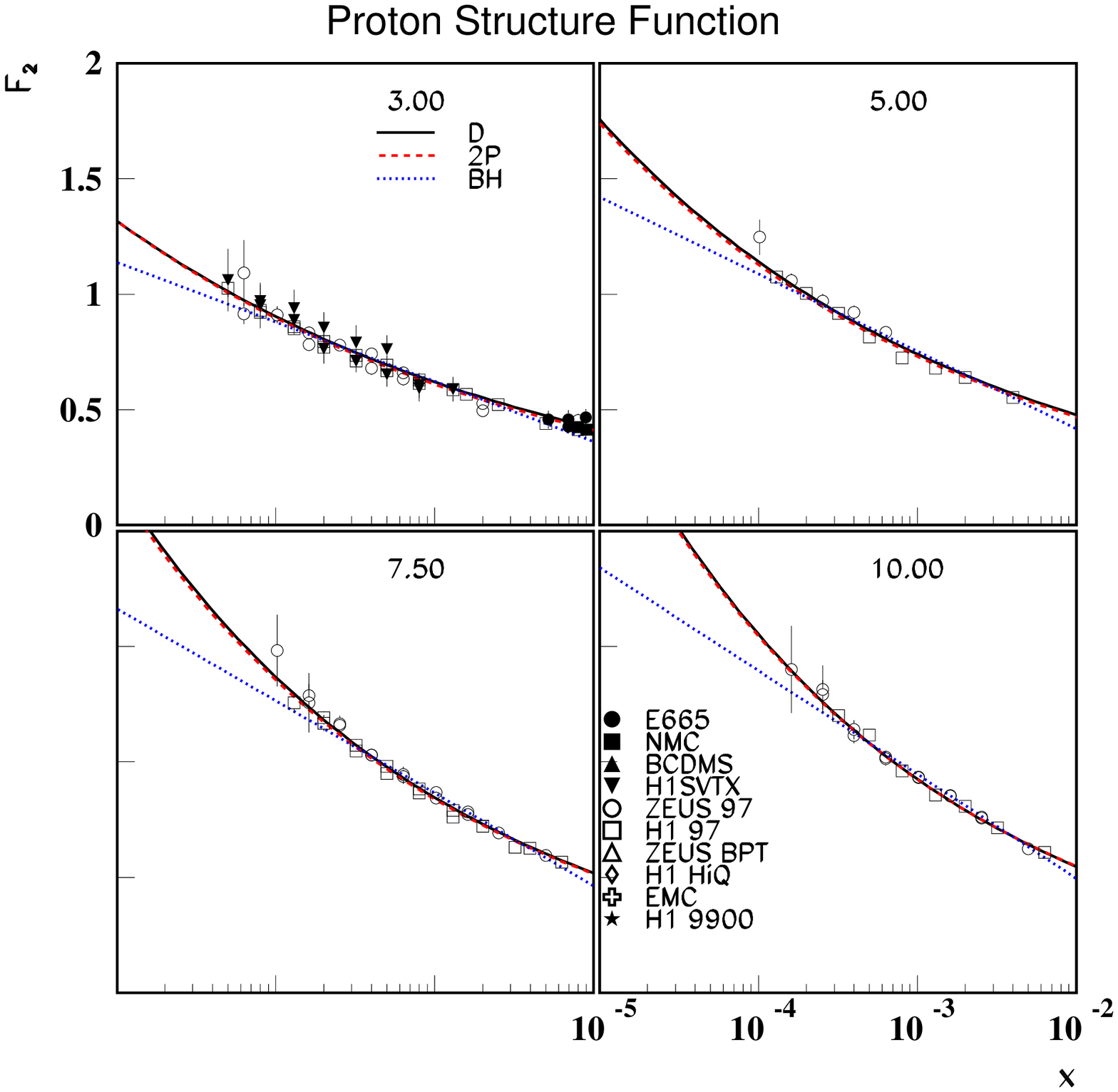}
\caption{\it See caption of Fig.~\ref{fig:f2center1}}
\label{fig:f2center2}
\end{center}
\end{figure}

\begin{figure}[hbpt]
\begin{center}
   \includegraphics[width=1.\textwidth]{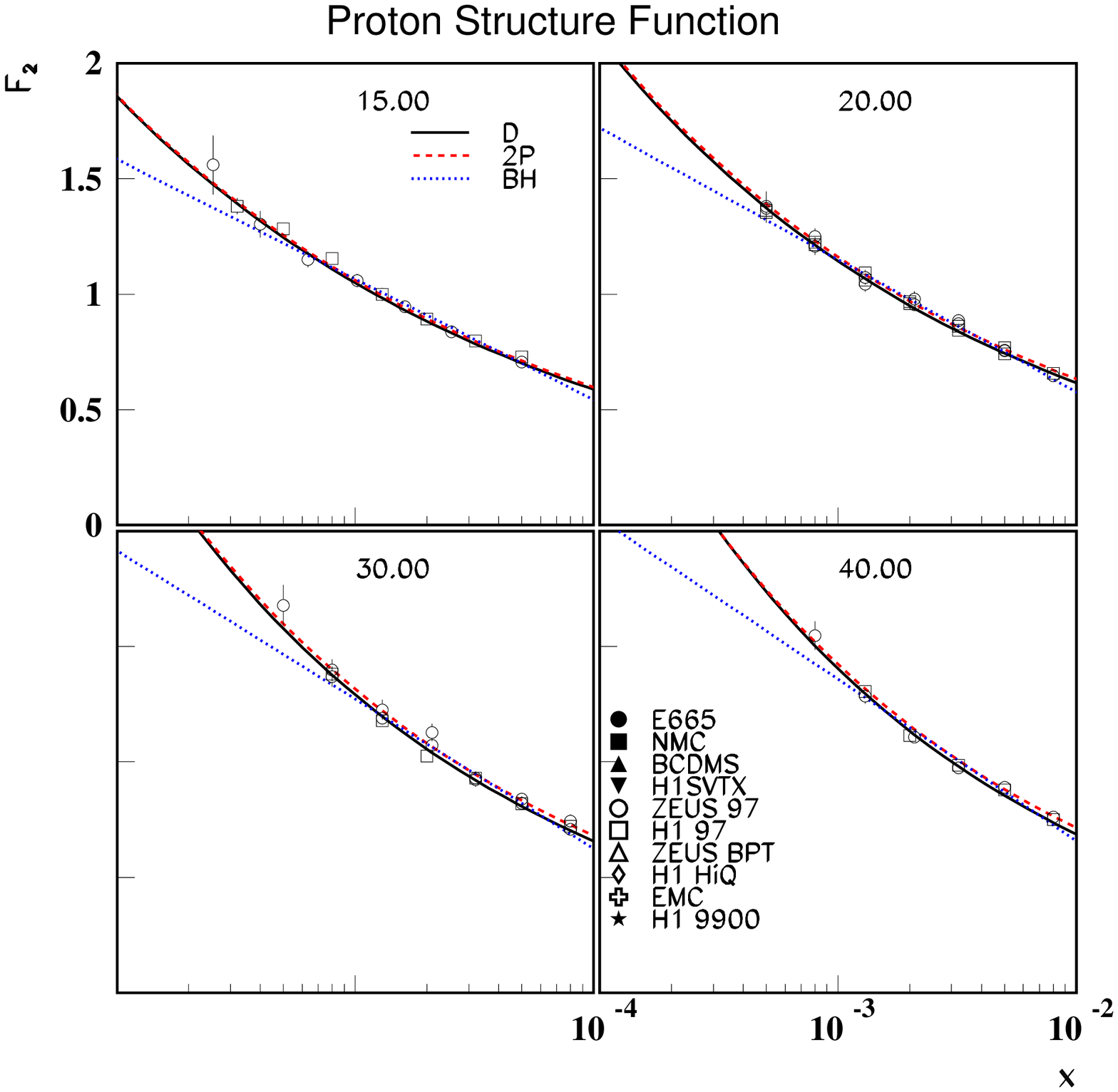}
\caption{\it See caption of Fig.~\ref{fig:f2center1}}
\label{fig:f2center3}
\end{center}
\end{figure}

\begin{figure}[hbpt]
\begin{center}
   \includegraphics[width=1.\textwidth]{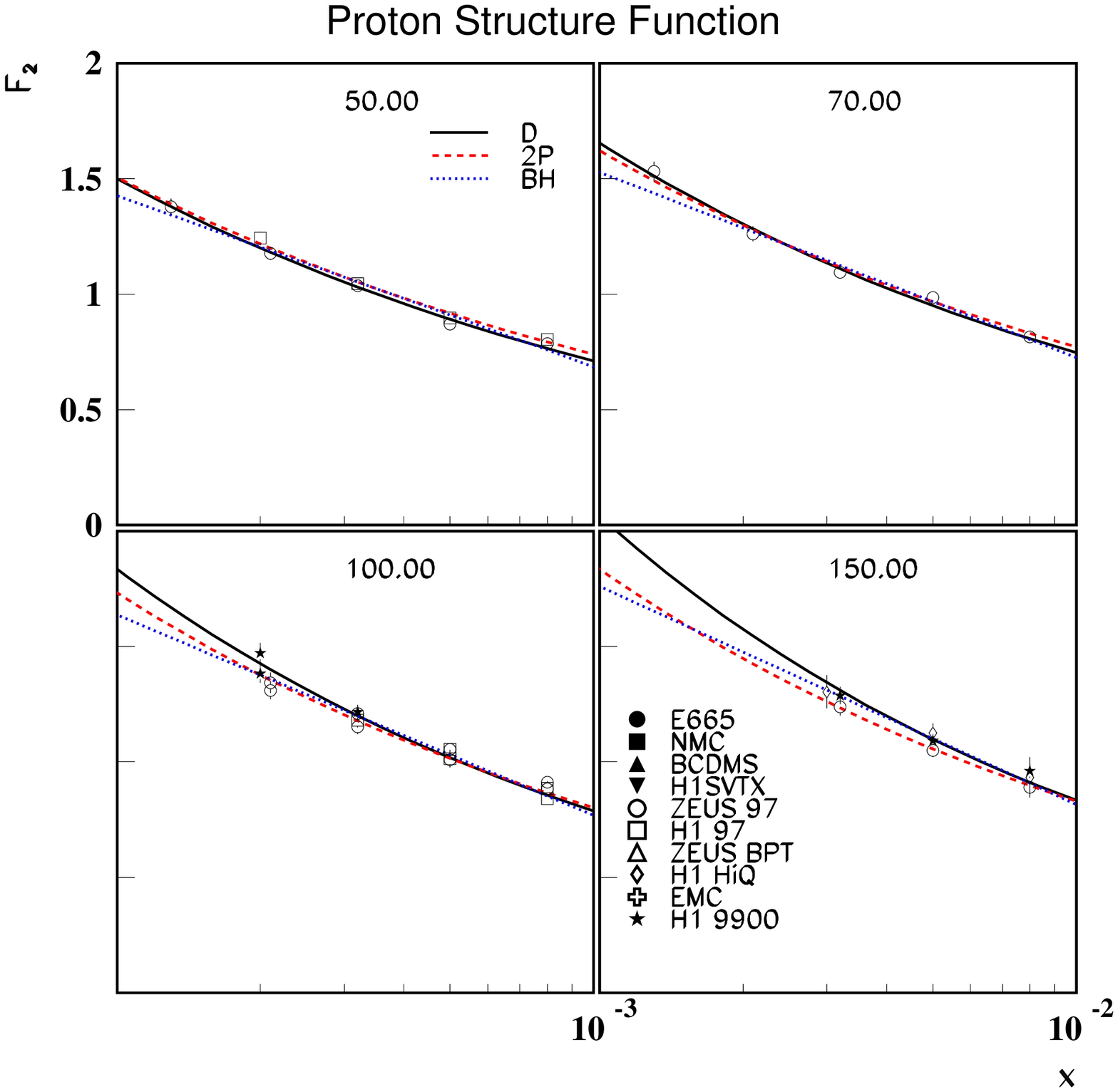}
\caption{\it See caption of Fig.~\ref{fig:f2center1}}
\label{fig:f2center4}
\end{center}
\end{figure}

\begin{figure}[hbpt]
\begin{center}
   \includegraphics[width=1.\textwidth]{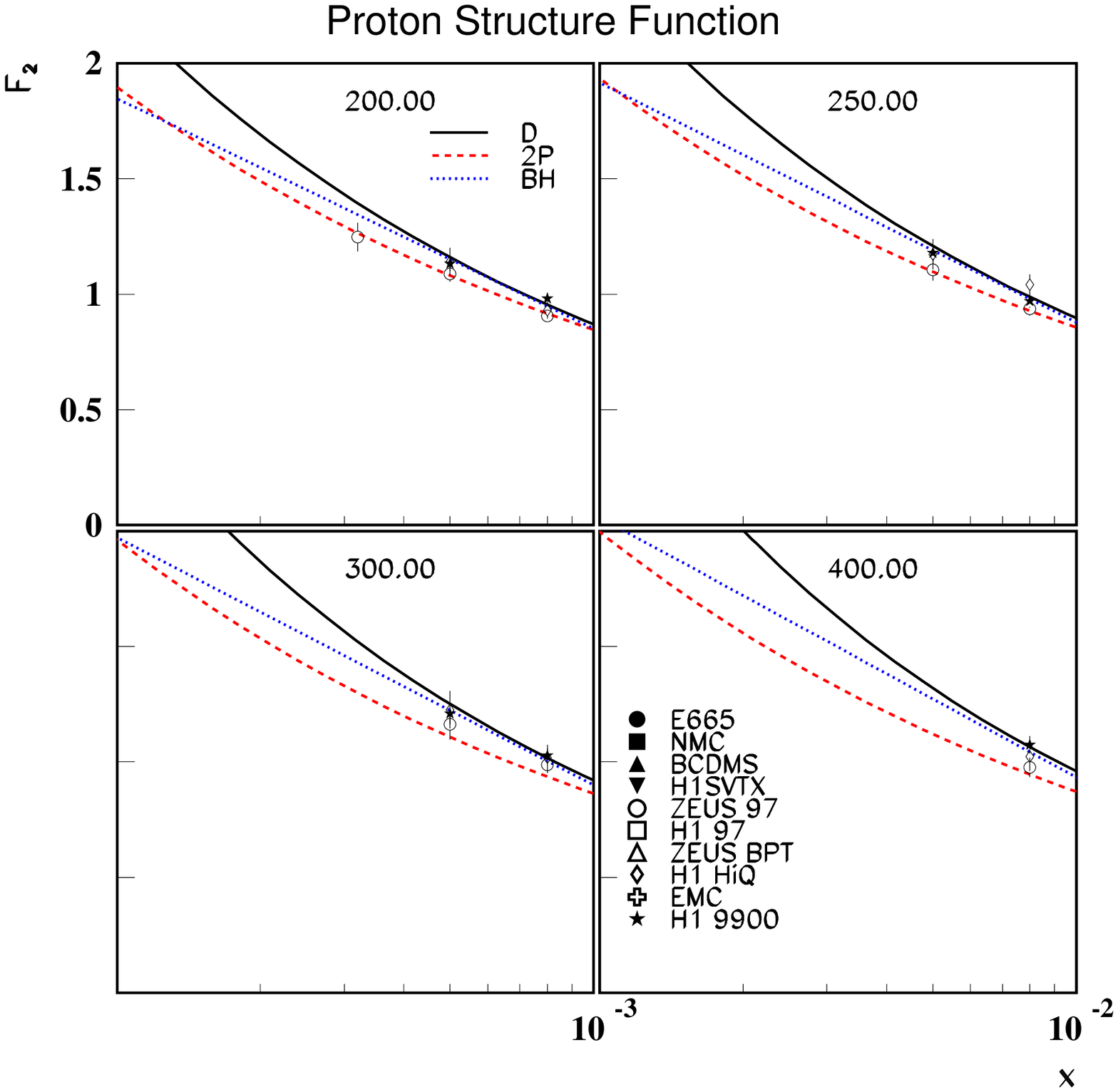}
\caption{\it See caption of Fig.~\ref{fig:f2center1}}
\label{fig:f2center5}
\end{center}
\end{figure}

The three parametrizations (fits 27, 28,29) are compared to bin-centered $F_2$ data in Figs.~\ref{fig:f2center1}-\ref{fig:f2center5}.  Fit 27 was used to bin center the data as explained in section~\ref{sec:center}. As can be seen from these plots, all parametrizations follow the general trend of the data.  However, the BH parametrization has a too-shallow $x$ dependence at low $Q^2$
while the D parametrization has a too-steep $x$-dependence at the highest $Q^2$. 

\subsection{Fits of parametrization D}
The value of $\sigma_0$ is the expected photoproduction cross section for $l=l_0$ and $Q^2=0$.  The value at larger $l$ is given by 
\begin{eqnarray*}
\sigma & = &\sigma_0 (l/l_0)^{\epsilon_0+\epsilon'\ln{Q^2_0}} \\
             & = &\sigma_0 (l/l_0)^{0.095} 
             \end{eqnarray*}
The exponent is within errors identical to the one found in \cite{ref:cudell} for hadron-hadron total cross sections.  In principle, this could be compared to measured photoproduction cross sections.  However, this would require knowledge of $l$ in photoproduction, whereas what is known is $W^2=Q^2/x$.  This comparison is therefore difficult.  However, it is possible to take the measured photoproduction cross section at HERA~\cite{ref:photoZEUS,ref:photoH1} of approximately $170$~{$\mu$}barn to deduce $l\approx 5\cdot 10^4$~fm; i.e., $x\approx 2 \cdot 10^{-6}$, similar to the $x$ range of the ZEUS BPT data. However, the photoproduction data are at smaller $Q^2$ and therefore should correspond to larger $l$, indicating possible problems with the parametrization as $Q^2 \rightarrow 0$.

The value of $M^2$ is within uncertainties the same as $M_{\rho}^2$, and the value of $l_0$ is very close to the proton radius.  The parametrization D has an interesting property, namely, that if $M=Q_0$, then the cross section will go to a fixed point at $l=l_0\exp{1/\epsilon'}$ independently of the starting $Q^2$.  In fact,
$$\frac{\partial \sigma^{\gamma P}}{\partial Q^2}|_{l}=0$$ requires 
$$\epsilon' \ln{l/l_0}=\frac{Q^2+Q^2_0}{Q^2+M^2} \;\;.$$
For $Q^2\rightarrow \infty$, this gives for fit 27 $l\approx 8\cdot 10^{6}~{\rm fm}$, while for $Q^2=0$ the value is
$l=5\cdot 10^{10}~{\rm fm}$.  A possible interpretation is that the data are headed for a fixed point, but the behavior softens once the evolved photon state approaches hadronic dimensions.  

\begin{figure}[hbpt]
\begin{center}
   \includegraphics[width=1.0\textwidth]{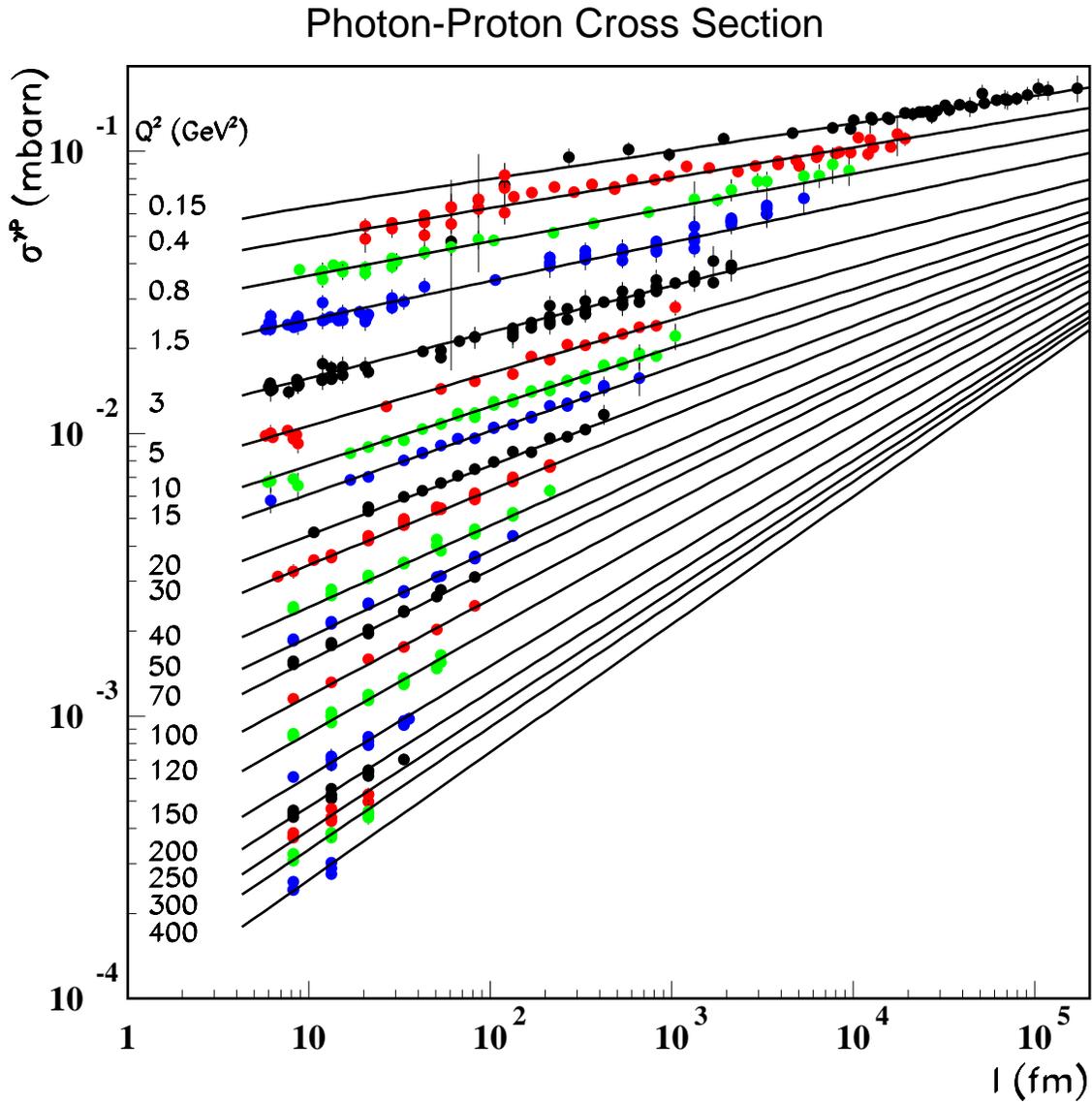}
\caption{\it The cross section $\sigma^{\gamma P}$ versus $l$.  The data have been bin centered using parametrization D and the fit values from fit 22.  The lines  are fitted curves of the form $\sigma=\sigma_0 l^{\lambda_{\rm eff}}$ for individual $Q^2$ values.}
\label{fig:sigtot}
\end{center}
\end{figure}

\begin{figure}[hbpt]
\begin{center}
   \includegraphics[width=1.0\textwidth]{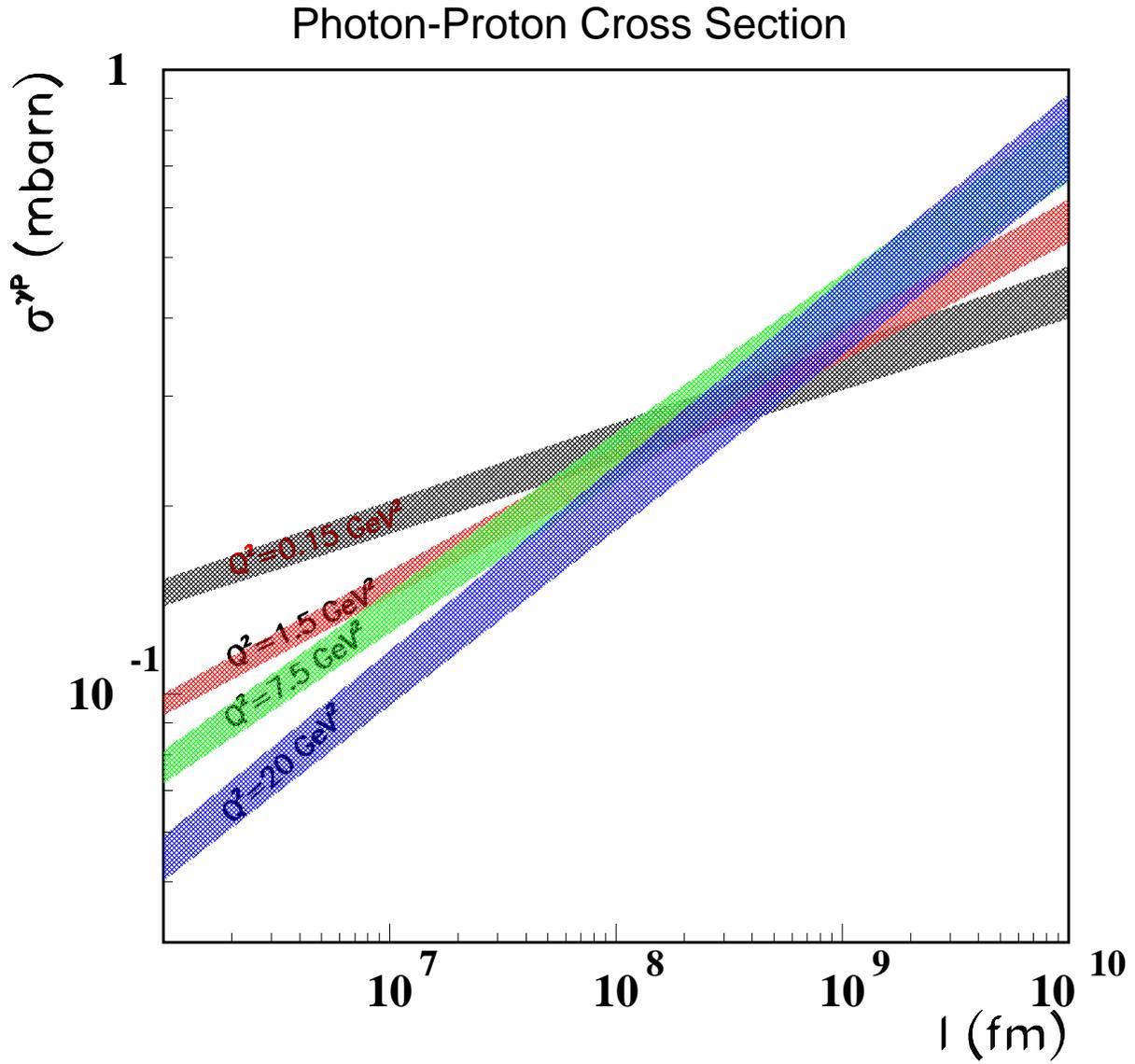}
\caption{\it An expanded view of the large $l$ region for the same conditions as those in Fig.~\ref{fig:sigtot}.  The 68~\% central interval from the fits are shown as bands.}
\label{fig:sigcross}
\end{center}
\end{figure}

The $\sigma^{\gamma P}$ cross section data is shown versus $l$ in Fig.~\ref{fig:sigtot}.  The data used here are bin centered using the values from fit 22, whereas the straight lines are from the fits to $\sigma=\sigma_0 l^{\lambda_{\rm eff}}$ for each $Q^2$ value.  A blow up of the large $l$ region, including the error bands from the fits, is shown in Fig.~\ref{fig:sigcross} for a subset of $Q^2$ values. 
The extrapolated $\sigma^{\gamma P}$ cross near $l=10^8$~fm.  At larger values of $Q^2$, the crossing point tends to move to lower $l$, but the uncertainties are larger.

\subsection{Fits of parametrization 2P}
Again here, the `soft-Pomeron' value found from hadron-hadron scattering is recovered within the uncertainties.  For $Q^2=0$, we find
\begin{eqnarray*}
\sigma & = &\sigma_0 (l/l_0)^{\epsilon_0} \\
             & = &\sigma_0 (l/l_0)^{0.086}
             \end{eqnarray*}
The value of the `hard-Pomeron' intercept is found to be $1+\epsilon_1\approx 1.37$, with equal contribution from the `soft' and `hard' Pomerons around $Q^2=\Lambda^2\approx 30$~GeV$^2$.  In this parametrization, there is no unique crossing point, and the dependence of $\sigma^{\gamma P}$ on $l$ becomes universal at high $Q^2$. 

Comparing to the measured photoproduction cross section at HERA as done above, we find a value of $l\approx 1.4\cdot 10^5$~fm for the HERA photoproduction data, which is somewhat larger than the result of the D parametrization.  A effective mean value of $l\approx 1\cdot 10^6$~fm is allowed within the parameter fit uncertainties and the uncertainties of the measured photoproduction cross section, which would correspond to the more reasonable effective mean $x\approx 10^{-7}$.  This again indicates that the 2P parametrization is a better representation of the data than the D parametrization.

\subsection{Fits of the parametrization BH}
As mentioned above, the fits with this extended version of the Buchm\"uller-Haidt parametrization does not work as well as the D or 2P fits.  The difficulty in the fitting is primarily at lower $Q^2$.  

The value of $M^2$ in these fits is considerably lower than in the D or 2P fits.  The reason is that a part of the softening of the $Q^2$ behavior is already accounted for in the logarithm.  The value of $Q^2_0$ is somewhat larger than the value found by Buchm\"uller and Haidt ($Q^2_0 =0.5$~GeV$^2$), while the value of $x_0$ is somewhat lower ($0.044$ versus $0.074$ in \cite{ref:BH}). 

\subsection{$\lambda_{\rm eff}$ as a function of $Q^2$}

One of the most striking results from the early HERA data was the dependence of the steepness of the rise of $F_2$ at small-$x$ on $Q^2$~\cite{ref:ZEUSpheno}.  At larger $Q^2$, a $\lambda_{eff} \propto \ln{Q^2}$ dependence was found, while the data indicated a flattening at smaller $Q^2$.  We use the data sets and fits described above to study the behavior in more detail.

\begin{table}[htdp]
\caption{\it Comparison of $\lambda_{\rm eff}$ values from the data using different parametrizations for the bin centering and different $x$ ranges.  The different fits are described in Table~\ref{tab:fits}.}
\begin{center}
\begin{tabular}{|c|cccccc|}
\hline
$Q^2$  &   &    &    &    &   &    \\
(GeV$^2$)  & 22 & 24 & 25 & 26 & 27 & 28 \\
\hline
$0.15$ & $0.098\pm0.007$ & $0.097\pm0.007$  & $0.102\pm0.007$ & $$ & 
$0.098\pm0.007$ & $0.104\pm0.007$ \\
$0.4$ & $0.097\pm0.003$ & $0.096\pm0.003$  & $0.102\pm0.003$ & $0.106\pm0.066$& $0.097\pm0.003$ & $0.102\pm0.003$  \\
$0.8$ & $0.120\pm0.005$ & $0.121\pm0.005$  & $0.130\pm0.005$ & $0.094\pm0.027$ & $0.123\pm0.005$ & $0.128\pm0.005$  \\
$1.5$& $0.142\pm0.003$ & $0.138\pm0.003$ & $0.144\pm0.003$ & $0.072\pm0.012$& $0.151\pm0.004$ & $0.154\pm0.004$ \\
$3$& $0.165\pm0.003$ & $0.162\pm0.002$ & $0.168\pm0.002$ & $0.151\pm0.008$& $0.169\pm0.004$ & $0.170\pm0.004$ \\
$5$& $0.193\pm0.004$ & $0.189\pm0.004$ & $0.194\pm0.004$ & $0.187\pm0.011$& $0.200\pm0.007$ & $0.200\pm0.007$\\
$7.5$& $0.217\pm0.004$ & $0.214\pm0.004$ & $0.216\pm0.004$ & $0.218\pm0.010$& $0.214\pm0.004$ & $0.214\pm0.005$ \\
$10$& $0.227\pm0.007$ & $0.227\pm0.006$ & $0.227\pm0.006$ & $0.231\pm0.015$& $0.228\pm0.007$ & $0.228\pm0.007$ \\
$15$& $0.247\pm0.006$ & $0.245\pm0.006$ & $0.245\pm0.006$ & $0.235\pm0.017$& $0.245\pm0.007$ & $0.247\pm0.007$\\
$20$& $0.259\pm0.005$ & $0.259\pm0.005$ & $0.260\pm0.005$ & $0.276\pm0.011$& $0.257\pm0.006$ & $0.258\pm0.006$ \\
$30$& $0.274\pm0.007$ & $0.275\pm0.006$ & $0.276\pm0.006$ & $0.272\pm0.010$& $0.277\pm0.008$ & $0.279\pm0.008$ \\
$40$& $0.289\pm0.008$ & $0.286\pm0.007$ & $0.289\pm0.007$ & $0.279\pm0.009$& $0.288\pm0.011$ & $0.290\pm0.010$ \\
$50$& $0.307\pm0.011$ & $0.303\pm0.009$ & $0.307\pm0.009$ & $0.305\pm0.009$& $0.314\pm0.014$ & $0.318\pm0.013$ \\
$70$& $0.323\pm0.014$ & $0.314\pm0.011$ & $0.313\pm0.010$ & $0.304\pm0.012$& $0.326\pm0.017$ & $0.328\pm0.018$ \\
$100$& $0.318\pm0.010$ & $0.318\pm0.009$ & $0.313\pm0.008$ & $0.313\pm0.007$& $0.317\pm0.014$ & $0.300\pm0.014$ \\
$150$& $0.328\pm0.022$ & $0.318\pm0.018$ & $0.307\pm0.016$ & $0.305\pm0.015$& $0.352\pm0.027$ & $0.343\pm0.029$\\
$200$& $0.329\pm0.024$ & $0.325\pm0.025$ & $0.324\pm0.018$ & $0.326\pm0.017$& $0.334\pm0.047$ & $0.346\pm0.047$\\
$250$& $0.327\pm0.033$ & $0.348\pm0.036$ & $0.340\pm0.020$ & $0.339\pm0.020$& $0.372\pm0.057$ & $0.336\pm0.059$ \\
$300$& $0.357\pm0.034$ & $0.372\pm0.041$ & $0.362\pm0.023$ & $0.364\pm0.024$& $0.322\pm0.067$ & $0.311\pm0.077$ \\
$400$& $0.306\pm0.066$ & $0.331\pm0.058$ & $0.352\pm0.035$ & $0.353\pm0.035$& $0.273\pm0.197$ & $0.137\pm0.090$ \\
\hline 
 \end{tabular}
\end{center}
\label{tab:lambdacomp}
\end{table}%

The quantity
$$\lambda_{\rm eff}\equiv \frac{\partial \ln \sigma^{\gamma P}}{\partial \ln l}|_{Q^2}$$
was calculated for each parametrization.  In terms of the parameters of the fits, we have
\begin{itemize}
\item[D:] $\lambda_{\rm eff}=\epsilon_0+\epsilon' \ln{(Q^2+Q^2_0)} $,
\item[2P:] $\lambda_{\rm eff}=\epsilon_0+(\epsilon_1-\epsilon_0) \sqrt{\frac{Q^2}{Q^2+\Lambda^2}} $,
\item[BH:] $\lambda_{\rm eff}=\frac{\ln{(Q^2/Q^2_0+1)}}{A+\ln{(Q^2/Q^2_0+1)}\ln{(l/l_0)}} $.
\end{itemize}

\begin{figure}[hbpt]
\begin{center}
   \includegraphics[width=1.0\textwidth]{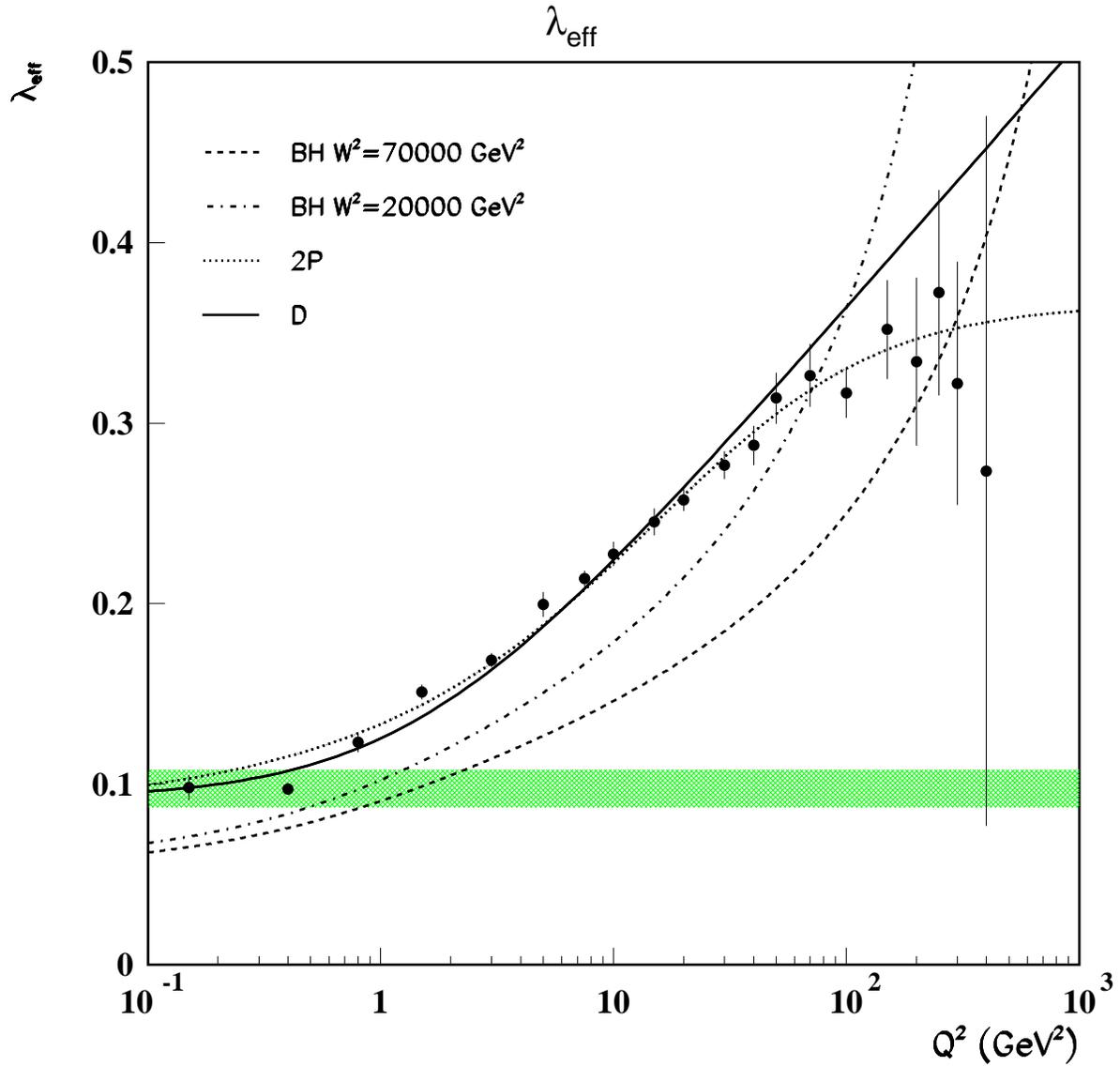}
\caption{\it The values of $\lambda_{\rm eff}$ as defined in the text as a function of $Q^2$ using centering from fit 27.  The dashed-dotted curve is for the BH parametrization (fit 29) and $W^2=20000$~GeV$^2$, while the dashed curve is for the BH parametrization with $W^2=70000$~GeV$^2$.  The solid curve is the D parametrization (fit 27), while the dotted curve is the 2P parametrization (fit 28).  The band represents the range of $\epsilon$ from fits to hadron-hadron scattering data~\cite{ref:cudell}.}
\label{fig:lambdaeff}
\end{center}
\end{figure}

The values of $\lambda_{\rm eff}$ for the different parametrizations are plotted versus $Q^2$ in Fig.~\ref{fig:lambdaeff} and a more complete set of results is given in Table~\ref{tab:lambdacomp}.  The value from the BH parametrization is $l$ dependent and therefore not given in the table.  Two curves are plotted in Fig.~\ref{fig:lambdaeff} for the BH parametrization, one for $W^2=20000$~GeV$^2$ and one for $W^2=70000$~GeV$^2$.  These values represent the typical range from HERA data.  The data in the plots used the bin centering from fit 27.

\begin{figure}[hbpt]
\begin{center}
   \includegraphics[width=1.0\textwidth]{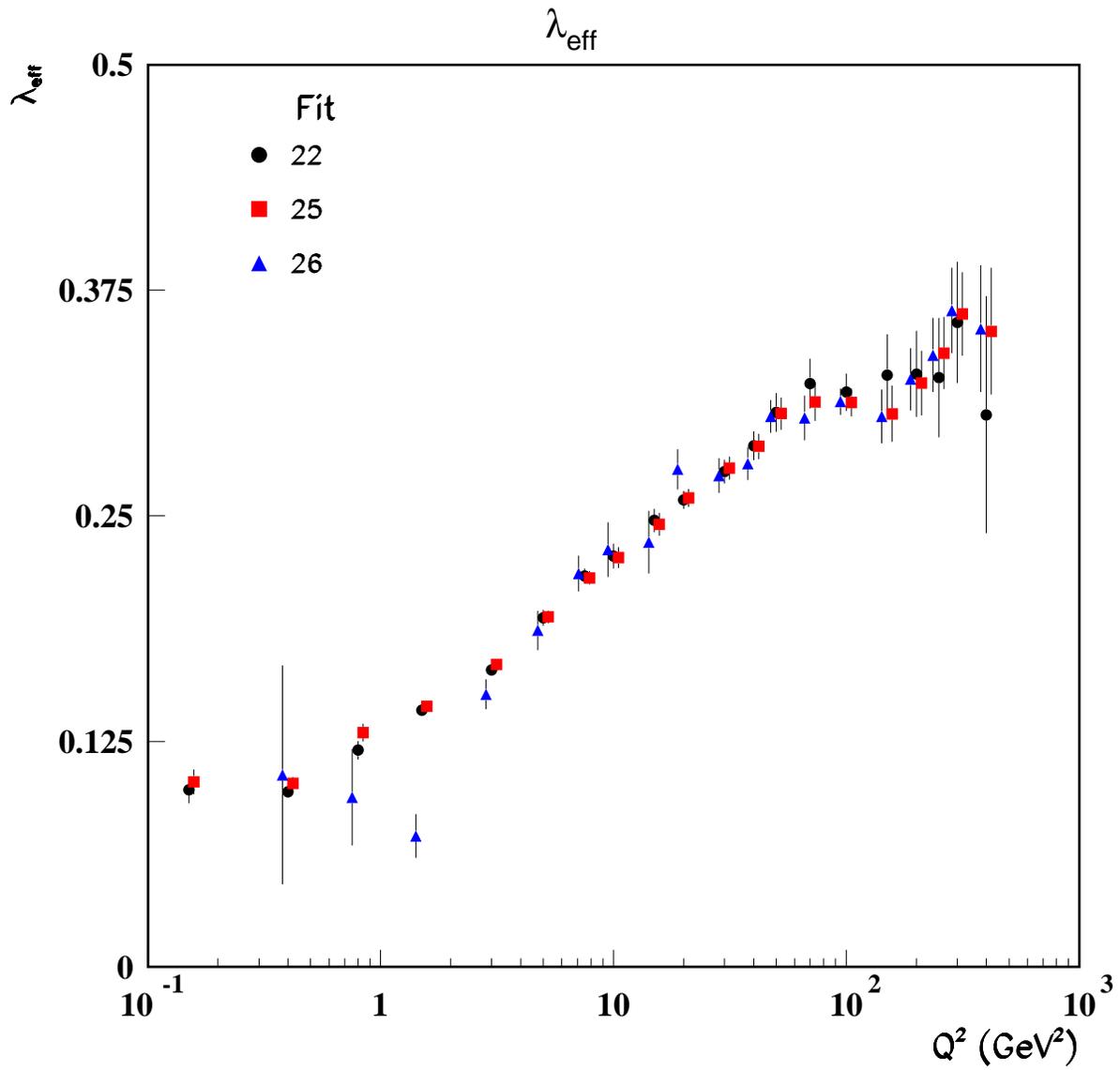}
\caption{\it Comparison of $\lambda_{\rm eff}$ data values for different kinematic ranges and different parametrizations as defined in Table~\ref{tab:fits}. The data points have been slightly offset along the $Q^2$ axis so as not to have overlapping uncertainty bands.}
\label{fig:lamdata}
\end{center}
\end{figure}

The D and 2P parametrizations follow the $l$ dependence of the data equally well up to $Q^2$ values of about $70$~GeV$^2$.  Beyond this value, the 2P parametrization is closer to the data.  Note that the mean value of $l$ for the existing data is decreasing as $Q^2$ increases.  It was checked whether the turnover of $\lambda_{\rm eff}$ at higher $Q^2$ is due to this by limiting the fitted values to $x>0.002$.  The results are presented in Table~\ref{tab:lambdacomp} (fit 26) and shown in Fig.~\ref{fig:lamdata}.  A turnover is still seen, although the uncertainties are of course much larger.   One point, at $Q^2=1.5$~GeV$^2$, shows a large discrepancy when the data is limited to the range $0.002<x<0.025$ (fit 26).  The reason is that with this cut, no HERA data are available at this $Q^2$.  The fitted uncertainty on the value of $\lambda_{\rm eff}$ is small because of the small uncertainties  of the fixed target data. 

It is difficult to conclude at this point how seriously to take the observed turnover of $\lambda_{\rm eff}$ at the higher $Q^2$ values.  While the effect seems to be real, it could still be explained by a combination of limited data and a changing $l$ range as $Q^2$ changes.  Note that only HERAI data have been used in this analysis.  Both H1 and ZEUS will report new high $y$ cross sections in the near future, which should greatly improve the measurement of $\lambda_{\rm eff}$ in this region, thus allowing for a better analysis in this kinematic region.

\section{Summary}
Parametrizations of $\sigma^{\gamma P}$ have been fitted to small-$x$ fixed target and HERA data using a Bayesian analysis based on a Markov Chain Monte Carlo.  Results for three different parametrizations:
\begin{eqnarray*}
{\rm D:} & \sigma^{\gamma P} & =  \sigma_0 \frac{M^2}{Q^2+M^2} \left( \frac{l}{l_0} \right)^{\epsilon_0+\epsilon'\ln{\left(Q^2+Q^2_0\right)}}  \\
 {\rm 2P:} &\sigma^{\gamma P} & =  \sigma_0 \frac{M^2}{Q^2+M^2} \left( \frac{l}{l_0} \right)^{\epsilon_0+(\epsilon_1-\epsilon_0)\sqrt{\frac{Q^2}{Q^2+\Lambda^2}}}  \\
{\rm BH:} & \sigma^{\gamma P} & =  \sigma_0 \frac{M^2}{Q^2+M^2} \left[A+\ln{\left(\frac{Q^2}{Q^2_0}+P_1\right)}\ln{\frac{x_0}{x}} \right]
                                   \end{eqnarray*}
                                   are given in this paper.  The best fits are from the 2P parametrization (inspired by the two-Pomeron model~\cite{ref:DL2}).  This conclusion goes hand-in-hand with the observation of a flattening of the effective growth of the cross section at large coherence length, $l$.  This flattening is intruiging, and should be verified with future HERA data. The D parametrization, inspired by the known properties of the HERA data, does not allow for this turn-over.  It also gives good fits overall, but clearly does not reproduce the observed $\lambda_{\rm eff}$ values at high $Q^2$.  The Buchm\"uller-Haidt~\cite{ref:BH} inspired parametrization, BH, does not follow the general trends of the data at the lowest and highest $Q^2$.
                                   
                                   A possibly interesting observation is the vanishing of the $Q^2$ variation of $\sigma^{\gamma P}$ when extrapolated to particular very large values of $l$.  This happens automatically in the D type parametrization at $l=l_0 \exp{1/\epsilon'}$ if $Q^2_0=M^2$. This behavior indicates that $\sigma^{\gamma P}$ will have a roughly $Q^2$ independent value for $l\geq 10^7$~fm.  An interpretation would be that the photon state has evolved sufficiently after this coherence length that it has lost memory of its initial configuration.  This speculation is based on extreme extrapolations, and more data would of course be needed to check this behavior.
                                   
	Many different parametrizations, not reported in this paper, have been attempted to fit the small-$x$ data, including some which gave very good fits.  They look very different in form, so that it is impossible to draw strong physics conclusions from the success of any one fit.  The possible turn-over of $\lambda_{\rm eff}$ at high $Q^2$ would provide a distinguishing feature which would rule out many possible forms and hopefully point to the correct physics.  Further, higher precision data from HERAII is expected on this front and could help clarify the situation.  More extensive and more precise data in the transition region at low $Q^2$ would also help in weeding out incorrect approaches.  Here, a proposed EIC~\cite{ref:EIC} would make a big difference.  For the extrapolations to very large $l$, obviously the LHeC~\cite{ref:LHeC} would be the best tool.

\section{Acknowledgments}
I would like to thank G\"unter Grindhammer, Henri Kowalski and Aharon Levy for many interesting discussions concerning the small-$x$ data, and Daniel Kollar and Kevin Kr\"oninger for fun and informative sessions on Bayesian data analysis.

\end{document}